\documentclass[sigconf]{acmart}
\renewcommand\footnotetextcopyrightpermission[1] 

\usepackage{float}
\usepackage[export]{adjustbox}
\usepackage{graphicx}
\usepackage{subcaption}
\usepackage[flushleft]{threeparttable}
\usepackage{xcolor}
\usepackage{makecell}
\usepackage{hyperref}
\usepackage{hhline}
\usepackage{comment}
\hypersetup{ 
    colorlinks=true,
    linkcolor=blue,
    filecolor=magenta,      
    urlcolor=blue,
    citecolor=blue
}

\usepackage{lipsum}%
\graphicspath{ {./images/} } 
\usepackage{tabularx}

\AtBeginDocument{%
  \providecommand\BibTeX{{%
    \normalfont B\kern-0.5em{\scshape i\kern-0.25em b}\kern-0.8em\TeX}}}
\begin{document}

\makeatletter
\def\thickhline{%
  \noalign{\ifnum0=`}\fi\hrule \@height \thickarrayrulewidth \futurelet
   \reserved@a\@xthickhline}
\def\@xthickhline{\ifx\reserved@a\thickhline
               \vskip\doublerulesep
               \vskip-\thickarrayrulewidth
             \fi
      \ifnum0=`{\fi}}

\newlength{\thickarrayrulewidth}
\setlength{\thickarrayrulewidth}{3\arrayrulewidth}

\pagestyle{plain} 

\title{Let's Vibrate with Vibration: Augmenting Structural Engineering with Low-Cost Vibration Sensing}

\author{Masfiqur Rahaman}
\authornote{Both authors contributed equally to this research.}
\email{1505111.mr@ugrad.cse.buet.ac.bd}
\affiliation{%
  \institution{Bangladesh University of Engineering \& Technology}
  \city{Dhaka}
  \country{Bangladesh}
}

\author{Nazmul Hasan Sakib}
\authornotemark[1]
\email{1505073.mnhs@ugrad.cse.buet.ac.bd}
\affiliation{%
  \institution{Bangladesh University of Engineering \& Technology}
  \city{Dhaka}
  \country{Bangladesh}
}

\author{Nafisa Islam}
\email{nafisaislam@pg.ce.buet.ac.bd}
\affiliation{%
  \institution{Bangladesh University of Engineering \& Technology}
  \city{Dhaka}
  \country{Bangladesh}
}

\author{Saiful Islam Salim}
\email{1018052067@grad.cse.buet.ac.bd}
\affiliation{%
  \institution{Bangladesh University of Engineering \& Technology}
  \city{Dhaka}
  \country{Bangladesh}
}

\author{Uday Kamal}
\email{udday2014@gmail.com}
\affiliation{%
  \institution{Bangladesh University of Engineering \& Technology}
  \city{Dhaka}
  \country{Bangladesh}
}

\author{Raihan Rasheed}
\email{1605062@ugrad.cse.buet.ac.bd}
\affiliation{%
  \institution{Bangladesh University of Engineering \& Technology}
  \city{Dhaka}
  \country{Bangladesh}
}

\author{A. B. M. Alim Al Islam}
\email{razi_bd@yahoo.com}
\affiliation{%
  \institution{Bangladesh University of Engineering \& Technology}
  \city{Dhaka}
  \country{Bangladesh}
}



\renewcommand{\shortauthors}{Masfiq and Nazmul, et al.,.}

\begin{abstract}
Using low-cost piezoelectric sensors to sense structural vibration exhibits a great potential in augmenting structural engineering, which is yet to be explored in the literature to the best of our knowledge. Examples of such unexplored augmentation include classifying diverse structures (such as building, flyover, foot over-bridge, etc.), and relating extents of vibration generated at different heights of a structure with the associated heights.
Accordingly, to explore these aspects, we develop a low-cost piezoelectric sensor based vibration sensing system aiming to remotely collect real vibration data from diversified civil structures. 
We dig into our collected sensed data to classify five different types of structures through rigorous statistical and machine learning based analyses. 
Our analyses achieve a classification accuracy of up to 97\% with an F1 score of 0.97. Nonetheless, in the rarely-explored \textit{time} domain, our analyses reveal a novel modality of relationship between extents of vibration generated at different heights of a structure and the associated heights, which was explored in the \textit{frequency} domain earlier in the literature with expensive sensors.
\end{abstract}


\maketitle

\section{Introduction}
In the domain of Structural Engineering, vibration pattern of civil structures exhibits applications in diversified areas such as designing architectures of the structures, Structural Health Monitoring, occupancy estimation, etc.,~\cite{goyal2016vibration, li2004full}. In the case of designing architectures, every structure follows specific vibration criteria that should be fulfilled when designing the structure. Concrete structures such as building, concrete foot overbridge, etc., are generally considered to generate less vibration. On the other hand, steel foot overbridge, suspension bridge, etc., generate more vibration. Besides, vibration of a structure varies at different heights.  These characteristics are very crucial when designing the architecture of a structure as misinterpretation of any of the characteristics or even ignorance of any of them may result in possible damage or structural health hazard in future.

\setlength{\parindent}{3ex} Due to the importance mentioned above, the dynamics of structural vibration has been investigated by several recent research studies~\cite{kuccukbay2017use, rivas2016road, sigmund2012analysis, richman2013seismic, garrityvibration, berlin2013sensor, magalhaes2012vibration, goyal2016vibration, li2004full}. However, these approaches lack some important considerations: (1) Most of the existing studies consider a single structure, i.e., bridge, building, railline, wind turbine, machine structures, etc.,~\cite{barbosa2016fault, joshuva2019selection, sigmund2012analysis, magalhaes2012vibration}. However, a study covering diverse civil structures is yet to be explored in literature to the best of our knowledge. 
(2) There exist research studies on implications of vibration generated by civil structures, e.g., structure and machine fault classification, engine classification, human identification, etc., from the pattern of vibration~\cite{ahmed2014automotive, barbosa2016fault, garrityvibration, kuccukbay2017use}. However, classifying diverse structures from their vibration patterns is yet to explored in literature. (3) There is a known relationship between different heights of a building and vibration generated at those heights. The relation only pertains to \textit{frequency} domain specially applicable for vibration data collected using high-cost sensors~\cite{li2004full}. It is yet to be explored how the relationship would be in the case of vibration data collected using low-cost sensors. Moreover, it is important to know whether the relationship in the case of vibration data collected using low-cost sensors would work in the conventional \textit{frequency} domain or it would get shifted to any other domain (such as the \textit{time} domain).

\setlength{\parindent}{3ex} Keeping all these considerations in mind, in this paper, we present a novel approach of classifying diversified civil structures based on their generated vibration. To do so, first, we devise and develop a low-cost piezoelectric vibration sensing module. Using the vibration sensing module, we build a diverse dataset of vibration sensed from five different types of civil structures through our year-long on-field data collection.
We show that there is a significant difference in vibration generated by different types of civil structures and the structures can be classified based on their generated vibration patterns. To the best of our knowledge, this finding is yet to be revealed in the literature. We also find a relationship between the mean \textit{amplitudes} of generated vibration and heights of different floors of a building. To the best of our knowledge, such a relationship persists only in the \textit{frequency} domain in the literature till now and the relationship is yet to be studied in the \textit{time} domain as done in this study.

\setlength{\parindent}{3ex} The overall methodology of our study includes developing a customized sensing system, deploying the sensing system in real settings, sending sensed vibration data to a server in the cloud and storing the vibration data there, showing the data on a real-time dashboard, performing statistical and machine learning based analyses for structure classification, and formulating relationship between floor heights and their associated vibrations. Here, for our machine learning based classification, we perform feature selection according to correlation and regression. Further, for Deep Learning based analysis, we perform hyperparameter tuning, i.e., tuning batch size, kernel size, number of filters, and choice of activation function. Besides, to find a relationship between the mean \textit{amplitudes} of vibration and corresponding heights of that floor, we perform linear curve fitting.


\begin{figure*}[!tbp]
    \centering
    \frame{\includegraphics[width=0.9\linewidth]{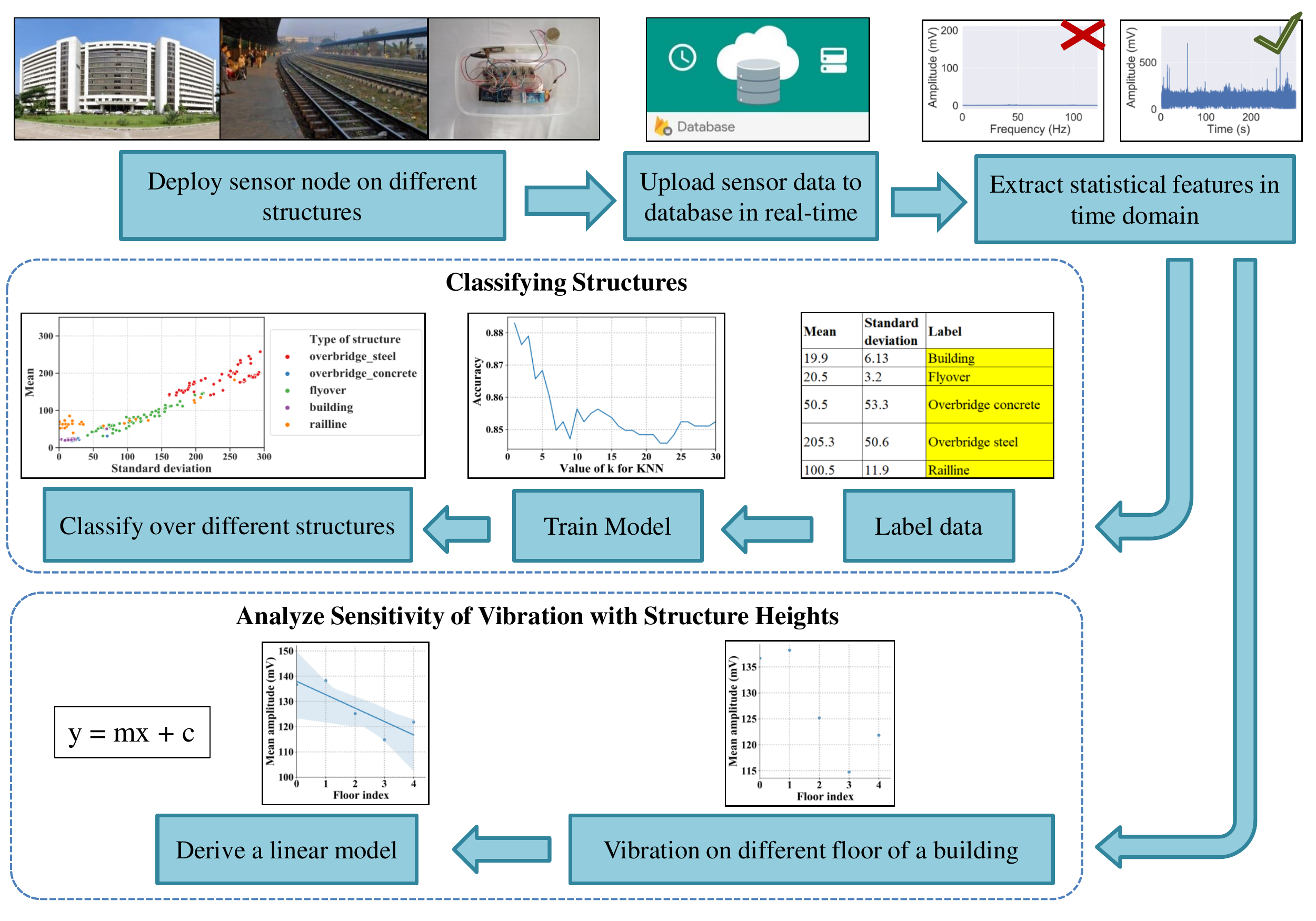}}
    \caption{High level illustration of our proposed methodology}
    \label{fig:methodology}
\end{figure*}

\setlength{\parindent}{3ex} Based on our study, we make the following set of contributions in this paper.
\begin{itemize}
    \item We design and develop a low-cost vibration sensing module using piezoelectric sensor. Using the sensing module, we collect real vibration data from 12 different civil structures having five different categories through a year-long on-field study.
    \item We classify the five different categories or types of civil structures based on their generated vibration through statistical and machine learning based analyses. Further, for achieving better accuracy, we develop a customized Deep Neural Network and utilize it for the classification task.
    \item Based on collected vibration data, we formulate a linear relationship between floor index of a building and vibration at that floor. The formulation explores \textit{time} domain analysis and considers both vertical and horizontal vibrations in formulating the relationship.
\end{itemize}

\section{Related Work and Our Motivation} 
In this section, we discuss existing studies in two directions: (1) vibration and its impact over diverse structures, 
(2) relating vibration with the heights of structures. Besides, we also discuss limitations of the existing studies, and thus, clarify our motivation behind this work.

\subsection{Vibration and Its Impact over Diversified Structures}
There have been several studies on detecting the source of vibration through different sensor-based data analytics \cite{berlin2013sensor, kuccukbay2017use, rivas2016road, sigmund2012analysis, richman2013seismic, garrityvibration}. For examples, Kucukbay et al.,~\cite{kuccukbay2017use} classified human, vehicle, and animal induced acoustic and vibration data. According to the type of vibration data, their proposed system triggers a camera event as an action for detecting intruders (human or vehicle). Besides, Rivas et al.,~\cite{rivas2016road} proposed a wireless sensor network on road that can precisely detect presences of vehicles. Their proposed system can calculate vehicle speed and travel direction from Accelerometer data. Sigmund et al.,~\cite{sigmund2012analysis} showed that vibration sensed from distant vehicles may be used to help in identifying key vehicle features such as engine type, engine speed, and the number of cylinders. 
Berlin et al.,~\cite{berlin2013sensor} classified train type and estimate train length from data accumulated by 3D MEMS Accelerometer. They studied Europe's busiest railroad sections and collects vibration patterns of 186 trains. They classified them into six categories using various methods. They also estimated the length of a train in wagons.

\begin{figure*}[!tbp]
    \frame{\includegraphics[width=.9\textwidth]{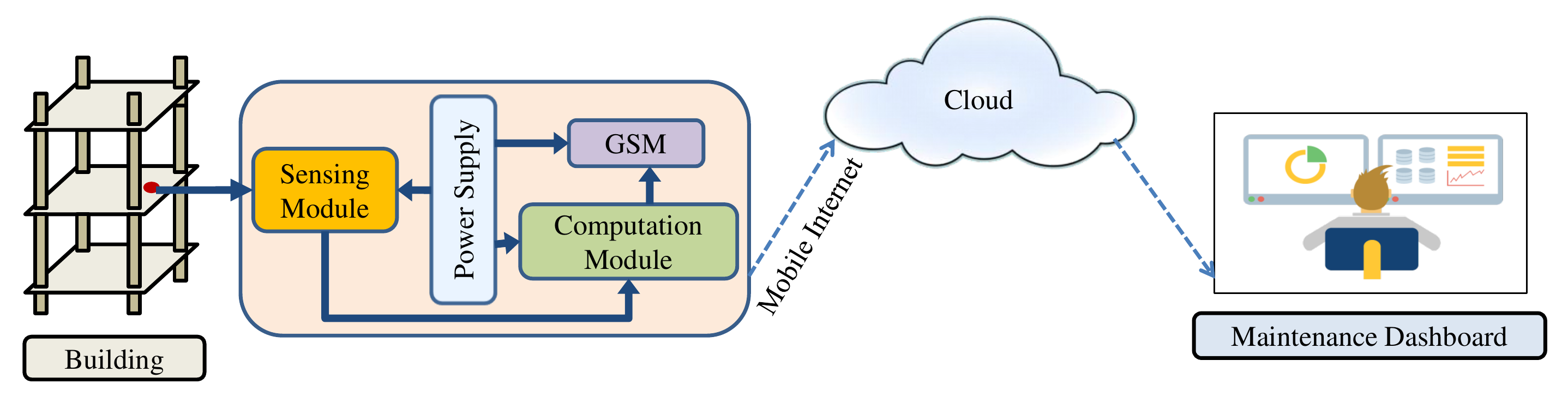}}
    \caption{System Architecture: sensing, communication, computation, power supply, and Maintenance Dashboard}
    \label{fig:system architecture}
\end{figure*}

\subsection{Change in Vibration with Varying Height}
There exist several studies regarding change in vibration with varying heights of a building ~\cite{pan2014empirical, michel2010full, li2004full, lagomarsino1993forecast, li2003effect, bindi2015seismic}. For examples, Pan et al.,~\cite{pan2014empirical} derived a relationship between the natural vibration periods and the height of high-rise public residential buildings in Singapore. They conducted ambient vibration tests on 116 buildings having a height ranging from 4 to 30 stories. The period-height relationships are derived using regression analysis. Besides, Li et al.,~\cite{li2004full} investigated an experimental and numerical study to investigate wind-induced vibrations and dynamic characteristics of a 63-story tall building. They analyzed the serviceability of this building under strong wind actions. It concludes from this study that the tall building will satisfactorily meet occupancy comfort criteria when it subjects to a strong typhoon with a generated wind speed.

\subsection{Vacancy in The Literature and Our Motivation}

\begin{figure}[!tbp]
    \centering
    \begin{subfigure}{0.22\textwidth}
        \frame{\includegraphics[width=\textwidth]{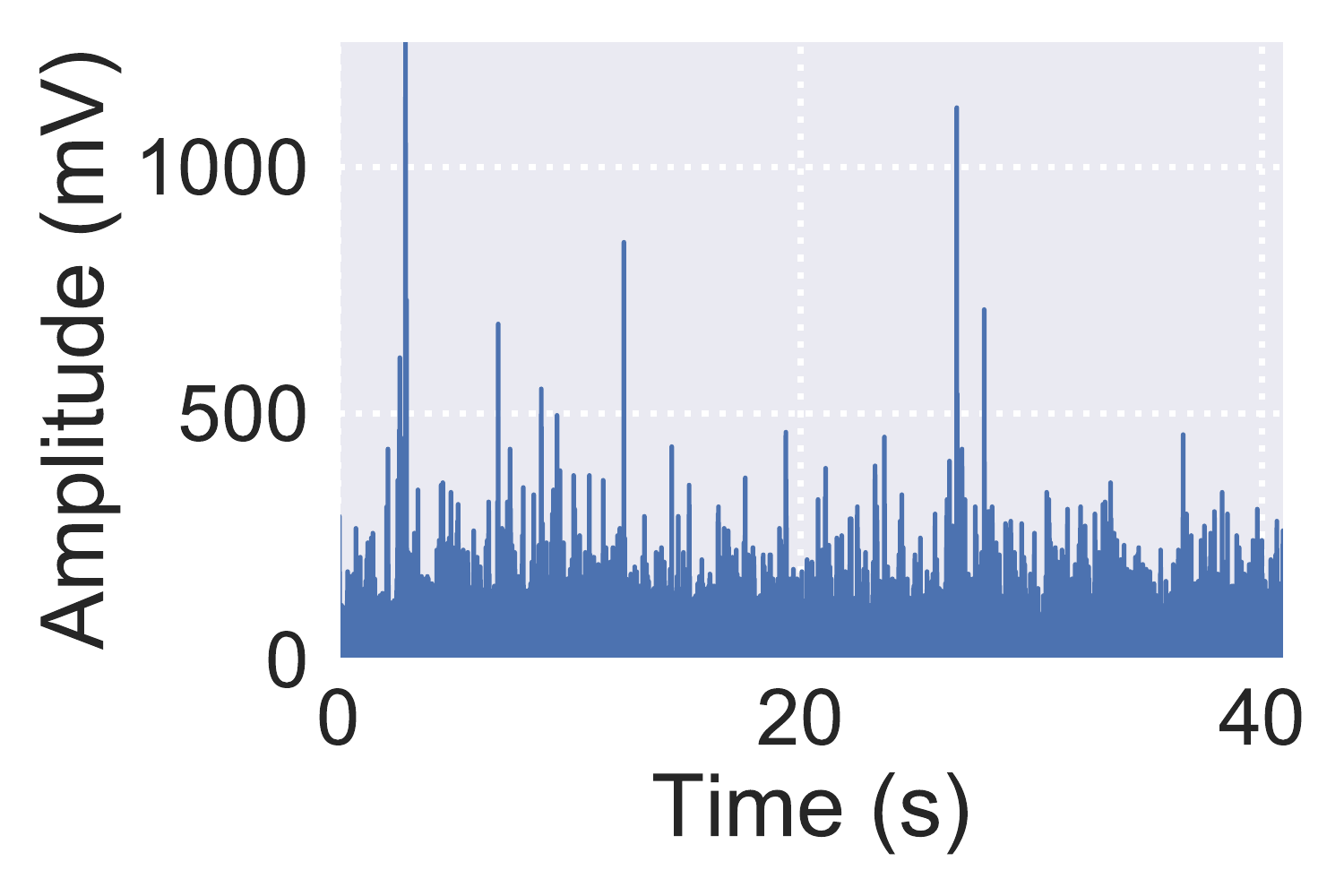}}
        \caption{Time domain}
        \label{fig:time}
    \end{subfigure}%
    \hfill
    \begin{subfigure}{0.22\textwidth}
        \frame{\includegraphics[width=\textwidth]{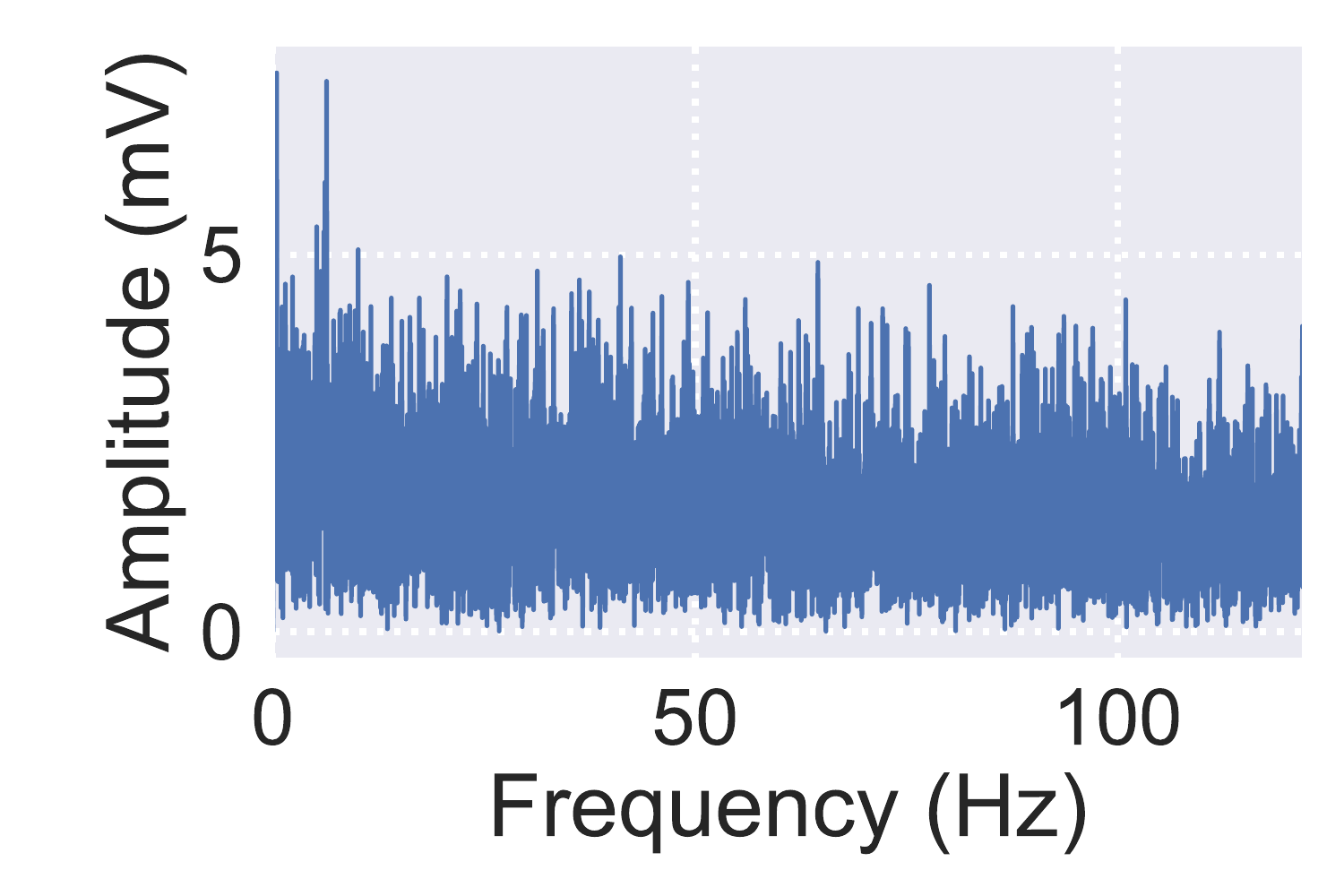}}
        \caption{Frequency domain}
        \label{fig:freq}
    \end{subfigure}%
        \caption{No significant frequency component after FFT}
        \label{fig:time_freq}
\end{figure}

\setlength{\parindent}{3ex} Among all the studies mentioned above, some studies focused on vehicle induced vibration, some focused on human-induced vibration for intrusion detection, some study related natural vibration period with height of structure, and so on.

However, classifying structures from their vibration characteristics is still unexplored in the literature. 
Nonetheless, existing studies focused on the \textit{frequency} domain. Thus, it has to be investigated whether we can still work on the \textit{frequency} domain or not while using the low-cost sensors. Finally, correlating vibration sensed by the low-cost sensor with the heights of building presents another perspective that is not explored in the literature.

To summarize, our contribution in classifying structures from their \textit{time} domain vibration may contribute in future in the field of SHM through classifying faults in structures. Also, this low-cost approach can correlate vibration with structure height which may be used in diagnosing faults considering height of structures. In this study we have not devised any SHM solution, rather proposed a new aspect of structure classification and correlated vibration with  height of the structure which may contribute to the \textit{time} domain signal processing in SHM.

\section{Methodology of Our Work}

In this section, we present how we classify different structures from vibration data. Then, we explain how vibration of different floors in a building varies with the heights of that floor. Figure~\ref{fig:methodology} shows a flow diagram of our proposed methodology. Here, First, we build the sensing module, and deploy it on the surfaces of building floors, flyover/overbridge spans, and rail-block. We collect and store vibration data to a database in real-time. Then, based on the collected vibration data, we formulate the problem of identifying source structure as a classification problem and attempt to solve it. Finally, we analyze the sensitivity of vibrations at different heights of a structure.

\subsection{Classifying Diversified Structures using Low-Cost Piezoelectric Sensors}
The most straightforward approach to classify structures involves directly performing FFT~\cite{fft} on the vibration signal, and then looking for the fundamental \textit{frequency} component for the signal. So, we first remove DC components from the time domain signal as shown in Figure~\ref{fig:time}.
However, as shown in Figure~\ref{fig:freq}, after FFT, we do not observe any obvious \textit{frequency} component. As a result, the classical approach of exploring fundamental frequency of the structure fails in the case of low-cost piezoelectric sensors. Also, we can not address the comparison between response of accelerometer and piezoelectric sensor as existing studies focused on the \textit{frequency} responses of the accelerometer and we can only analyze time domain response of piezoelectric sensors~\cite{goyal2016vibration, li2004full, magalhaes2012vibration}. 

\setlength{\parindent}{3ex} Hence, we move forward to time-domain analysis. In time domain, we first extract different statistical features from raw data points. We label the data according to their sources. In our study, we explore five different sources covering building, flyover, steel overbridge, concrete overbridge, and railline. We collect vibration data from 12 different locations in Dhaka city covering the above mentioned five classes of structures. For flyover, we collect data for five different spans. In the case of building, we collect data from every floor. Subsequently, we train several classification models to classify each type of structure. We present detail results and findings in Section~\ref{dimensionality} and \ref{deep learning}.


\subsection{Analyzing Sensitivity of Vibration at Different Heights of A Structure}

In our study on classifying structures, we consider a 11 storey building at Bangladesh University of Engineering and Technology campus in Dhaka city. Here, we take vibration data from each of the floors. We observe that each floor poses different ambient vibration, and we find a linear relationship between mean of the time series floor vibration and the height of the associated floor.

\setlength{\parindent}{3ex} For further investigation, later, we study 4 multi-storey buildings in Dhaka city for analyzing vibration characteristics with varying floor heights. Here, we collect floor vibration data for each floor of the building using the piezoelectric sensing module. We deploy the piezoelectric disc on the surface of the floor. Then, for each floor, we take ambient vibration data. Based on the captured data and our subsequent analysis, we find a linear relationship between the mean \textit{amplitude} of floor vibration and floor height as stated earlier. We present detail results and findings in Section~\ref{section: vibration and building height}.


\section{Customized Sensing System}
\label{setup1}

\begin{figure}[!tbp]
    \frame{\includegraphics[width=.9\linewidth]{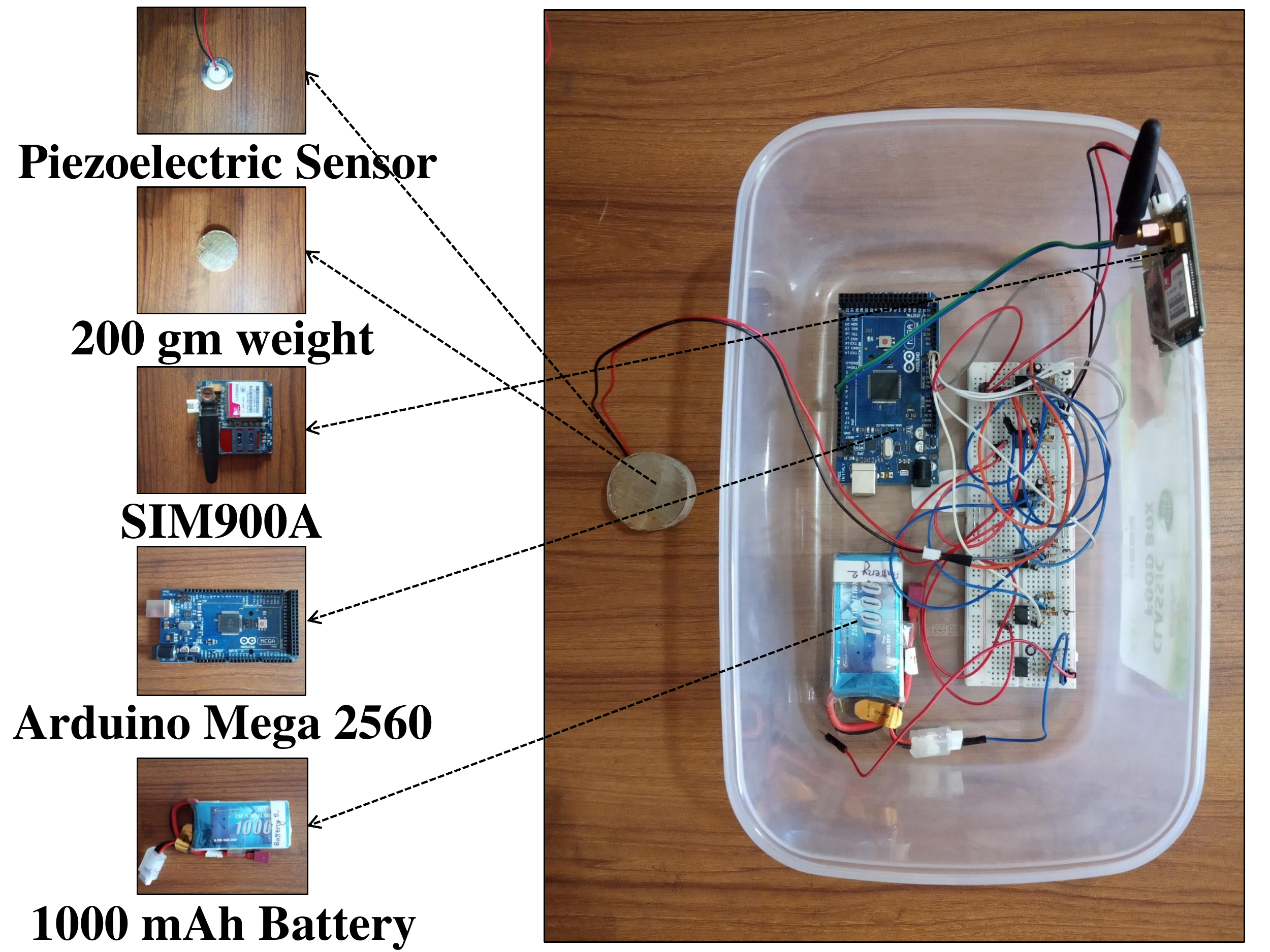}}
    \caption{Our hardware setup. We use 200 gm weight bar on top of piezo sensor to fix the piezo disk and get stable signal.}
    \label{fig:setup}
\end{figure}

To perform our intended tasks, we design and develop a customized sensing system. Here, we use low-cost components to make sure that the whole system remains low-cost in nature. Main components of our system include sensing module, computational module, communication module, power supply module, and a real-time dashboard. Figure~\ref{fig:system architecture} shows the architecture of our proposed system and Figure~\ref{fig:setup} shows the full system. We elaborate each of the components below.

\subsection{Sensing Module}

\begin{figure}[!tbp]
\centering
    \frame{\includegraphics[width=.9\linewidth]{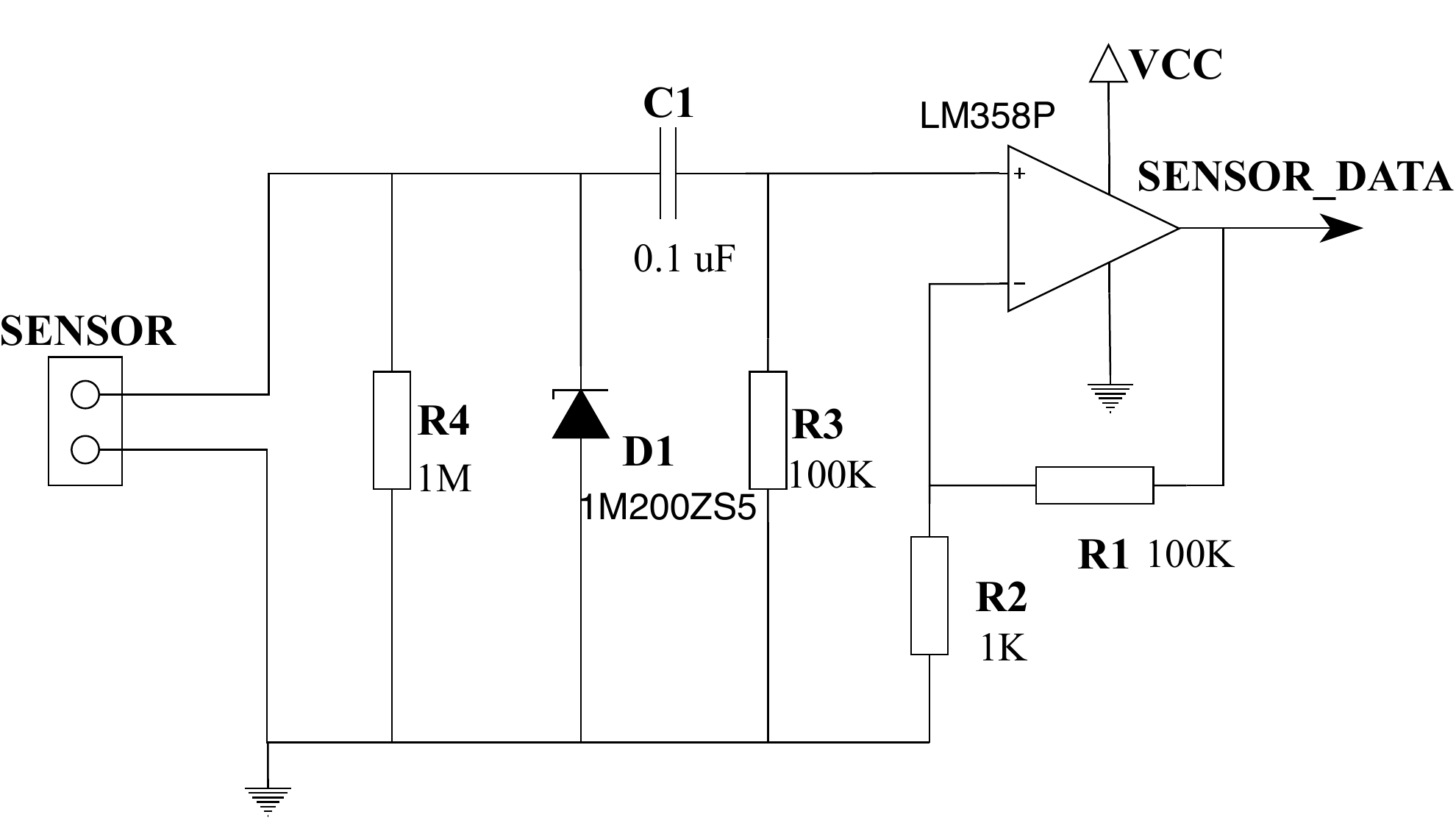}}
    \caption{Circuit diagram of Sensing module}
    \label{fig:sm}
\end{figure}

We use a low-cost piezoelectric disc~\cite{piezo} to sense the ambient vibration of structures. The raw analog signal collected by piezoelectric disc is first amplified through an amplification circuit as shown in Figure~\ref{fig:sm}. We choose the LM358P as the operational amplifier. LM358P is a low power dual operational amplifier. The reason behind choosing LM358P is explained here. The natural \textit{frequency} of bridge and other concrete-made structures varies in range 2-4 Hz , but values 0-14 Hz have also been reported~\cite{bachmann2012vibration}. LM358P's cutoff frequency is 200 Hz which is favorable considering the input signals' frequency response. The amplification factor of the whole amplifier circuit is 100. We have also tried amplification factor of 200, 500, and 1000. However, for some structural vibration, signal cuts at ADC value 1023 for amplification factor greater than 100. Also, more the amplification factor, more the power consumption. That is why we choose the amplification factor of 100. Then, we feed the amplified signal to a 10-bit analog-to-digital converter (ADC) on an Arduino Mega~\cite{arduino_mega} whose range is 0 to 5 V having the maximum sampling frequency of 9615 Hz which is greater than op-amp's output signal (200 Hz). Thus, the choice of LM358P supports both the input signals' frequency response and the sampling frequency of ADC.

\subsection{Computational Module}
We use the Arduino Mega 2560 as our computational module. It takes sensed data from the sensing module at an interval of eight seconds. Subsequently, it determines 12 statistical features from the captured time-series data. The statistical features are mean, median, mode, standard deviation, max, min, rms, total number of peaks, average of peak values, skewness, kurtosis, and creast factor~\cite{scipy, numpy}.

\subsection{Communication Module}
We use SIM900A~\cite{sim900a}, which is a GSM-based device, to send statistics of our sensed data to the server in real time. The use of SIM900A gives robustness to our system by providing network support outside home or office where WiFi or broadband is not available. This is why we can deploy our sensing module at diversified places covering the structures, such as flyovers, overbridges, and raillines.

\subsection{Power Supply Module}
We can choose either direct or battery power options as a source of power. In room environment, we use a direct power supply unit with a 220 V to 5 V adapter. On the other hand, for outdoor cases such as flyover and overbridges, we use a 1000 mAh battery as the source of power.

\subsection{Real-time Dashboard}
We send the statistical measurements as HTTP post request in a URL which is then stored in a database.  We develop a dashboard to display the data points in real-time. The locations of deployed nodes and their status can be found here (kept hidden for the purpose of anonymity).








\subsection{Cost Analysis of Proposed System}

For large-scale deployment of any system, costing of the whole system is very important. This is why, we try to minimize the cost of our proposed system to make it a low-cost solution. Table~\ref{tab:pricetab} presents a breakdown of equipment costing of our system. Equipment cost of the system is \~50 USD per unit, which is comparable to that of a widely-adopted smartphone unit. Thus, our system exhibits a potential to be a ubiquitous solution.

\begin{table}[!tbp]

\caption{Costing of necessary hardware equipment}
\label{tab:pricetab}

\resizebox{\linewidth}{!}{
\begin{tabular}{ |c|c|c|c|}
\hline
Component name            & Model name          & Quantity  & \begin{tabular} {@{}c@{}}Unit price (USD)
\end{tabular}\\ \hline

{Piezoelectric sensor} & {7BB-20-6L0}   & {1}       & {1}                      \\ \hline
{Amplifier}        & {LM358P}            & {1}       & {6}                      \\ \hline
{Resistor}         & {1M,1K,100K}       & {4}       & {0.25}                       \\ \hline
Zener Diode & 5 V & 1 & 0.5 \\ \hline
Capacitor & 0.1 uF & 1 & 0.5 \\ \hline
Microcontroller & Arduino Mega2560 & 1 & 14 \\ \hline
GSM Module & SIM900a & 1 & 16 \\ \hline
Power supply & \begin{tabular}{@{}c@{}}Polymer Lithium Ion \\ Battery - 1000mAh\end{tabular} & 1 & 10 \\
\hline
{Total price}               & \multicolumn{3}{c|}{{50 USD per node}} \\ \hline
\end{tabular}
}

\end{table}

\section{Deployment and Preparation of Dataset}
We deploy our sensing module in such a way that it retains the flavor of diversity. We collect data from the real deployments to prepare our dataset.

\subsection{Deployment over Diversified Structures}
\label{section:deployment}

\begin{figure}[!tbp]
    \frame{\includegraphics[width=.8\linewidth]{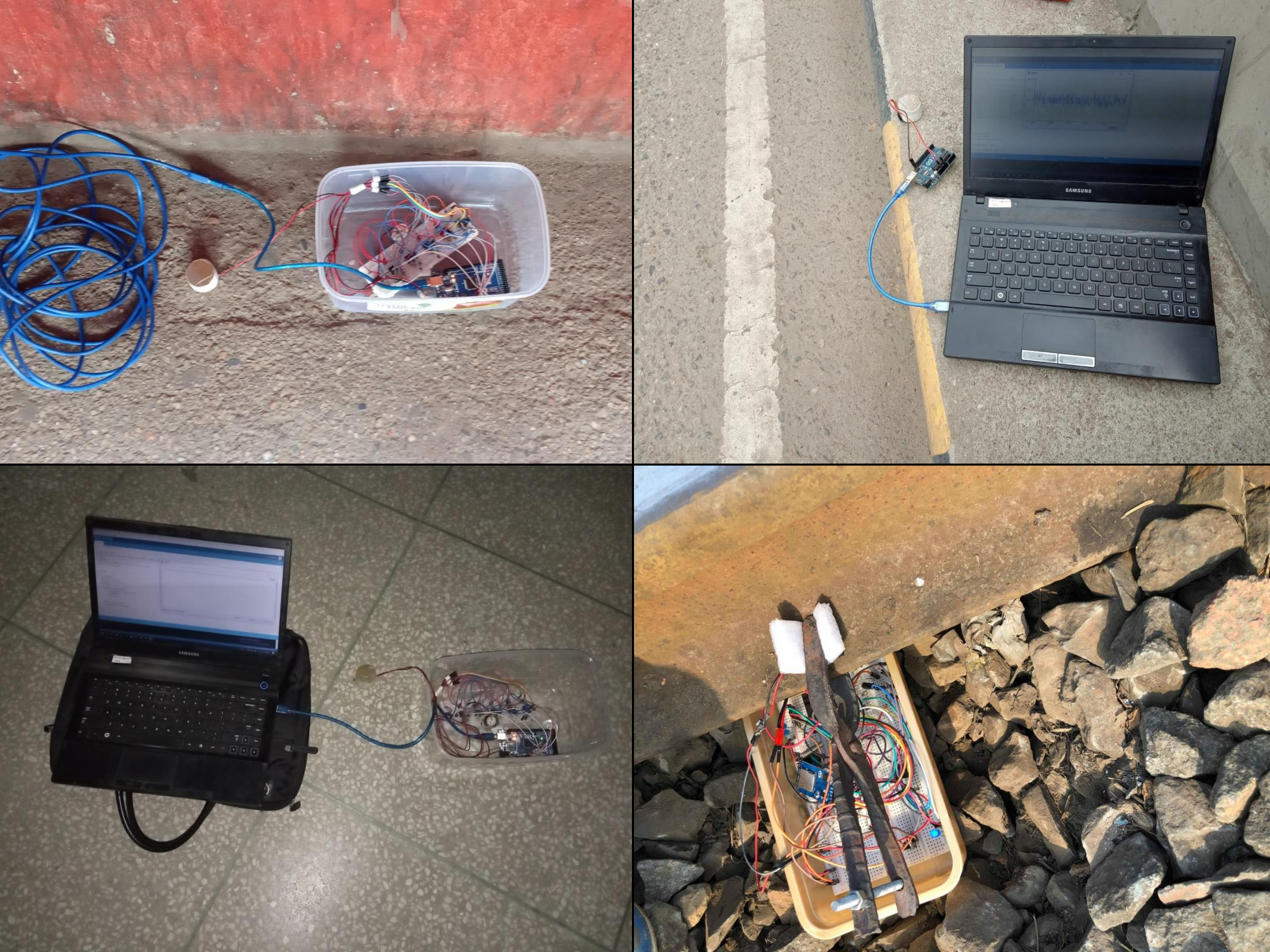}}
    \caption{Deploymentment of sensor nodes on different structures: Overbridge, Flyover, Building, and Railline}
    \label{all deployments}
\end{figure}

\begin{figure*}[!tbp]
\centering
    \begin{subfigure}[t]{0.5\textwidth}
        \centering
        \frame{\includegraphics[width=.8\linewidth]{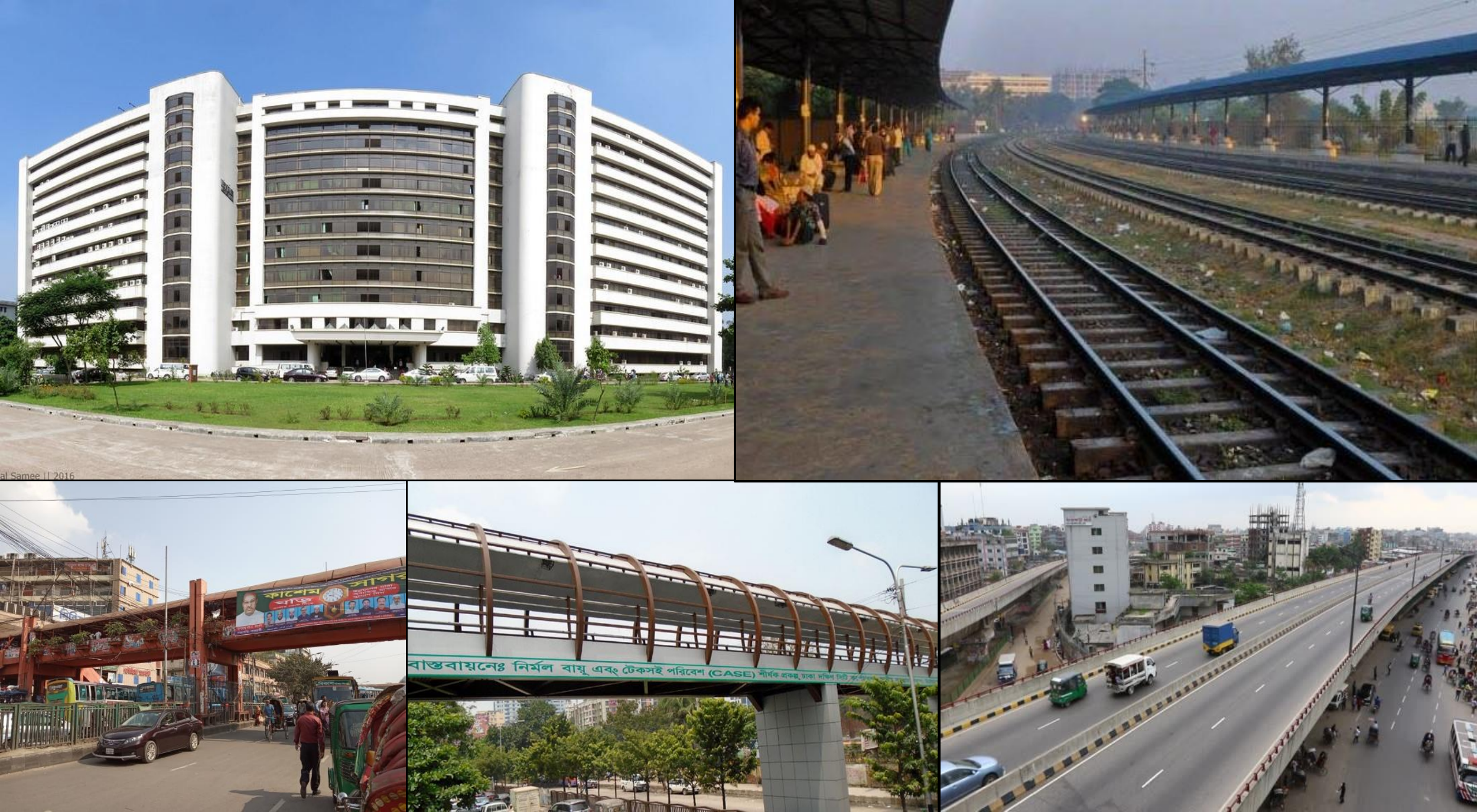}}
        \caption{Building, railline, concrete overbridge, steel overbridge, flyover}
        \label{5 structures}
    \end{subfigure}
    \begin{subfigure}[t]{0.48\textwidth}
        \centering
        \frame{\includegraphics[width=.8\linewidth]{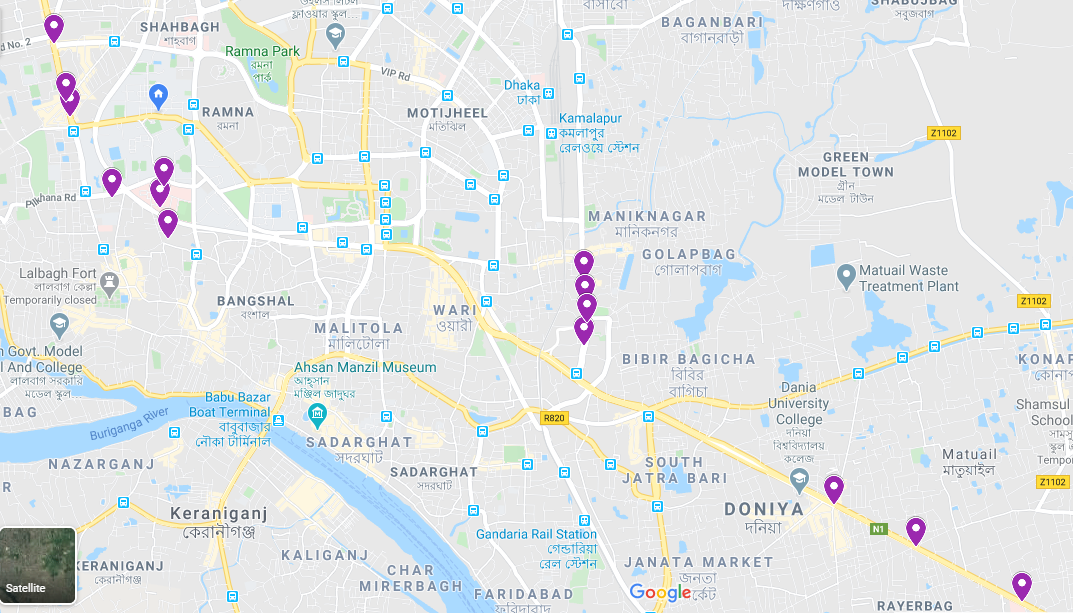}}
        \caption{12 locations pinned where sensor node is deployed }
        \label{locations}
    \end{subfigure}
\caption{Overview of data collection subjects and locations}
\label{data collection}
\end{figure*}

We deploy our sensing module in 12 different locations in Dhaka city shown in Figure~\ref{locations}. This enables sensing from 12 structures having five different categories among them as shown in Table~\ref{flyoverinfo}, \ref{buildinginfo}, \ref{raillineinfo}, \ref{steeloverbridgeinfo}, and \ref{concreteoverbridgeinfo}. The five different categories are flyover, building, steel overbridge, concrete overbridge, and railline. In all cases, we place the sensor on a horizontal surface to sense vertical vibration. Figure~\ref{all deployments} shows some snapshots of such deployment.


\subsection{Preparation of Data-set}

As mentioned earlier a web server keeps all data collected by the sensor. We organize the data by location and type of structure and store them accordingly. From all structures under investigation, we collect data of a total interval of 4 hours and 20 minutes. As we have on an average 200 data points at each second, our data set contains summary of a total of 3 million raw data points. To be specific, our dataset contains 1,159 summary data points.

\setlength{\parindent}{3ex} In the case of flyover (Table~\ref{flyoverinfo}), we consider four different spans for data collection. We choose the middle of each span to deploy our sensing module so that maximum vibration can be captured. Besides, we deploy the module on both left and right sides of the flyover to achieve symmetry as well as diversity.

\setlength{\parindent}{3ex} In the case of foot overbridges (Table ~\ref{steeloverbridgeinfo}, and ~\ref{concreteoverbridgeinfo}), Our data set contains data collected from four different steel-made and one concrete-made foot overbridges. When we collect data from these structures, varying number of crowds: light, medium, dense are crossing over the bridges. We collect data for at least 2 different positions on each overbridge.

\setlength{\parindent}{3ex} We collect data from four academic and residential buildings (Table ~\ref{buildinginfo}). In each building, data from every floor contribute to our data set. In case of two of the buildings, vibration of two pillars contribute the data set.

\setlength{\parindent}{3ex} We also cover raillines (Table ~\ref{raillineinfo}). In raillines, data from both meter gauge and broad gauge lines, contribute to the data set. Here, the data is collected only when no train passes by. We cover crossings over railline where buses, cars, bikes, cycles, and people cross railline from one side to another.

\setlength{\parindent}{3ex} After collecting the time-series data, we extract 12 statistical features from those data points. Table~\ref{sampledataset}. shows some sample entries in our dataset.

\begin{table*}[!tbp]
\centering
\caption{Details of flyover where sensor module is deployed}
\label{flyoverinfo}
\begin{tabular}{|c|c|c|c|} \hline
Flyover & \#Spans covered & Duration of data collection (minutes) & Position of deployment\\ \hline
Flyover-1 & 5 & 30 & on surface of road \\ \hline
\end{tabular}
\end{table*}

\begin{table*}[!tbp]
\centering
\caption{Details of building where sensor module is deployed}
\label{buildinginfo}
\begin{tabular}{|c|c|c|c|c|} \hline
Building & Type of building &\#Floors & Duration of data collection (minutes) & Position of deployment\\ \hline
Building-1 & Office & 11 & 60 & floor, pillar \\ \hline
Building-2 & Office & 5 & 30 & floor, pillar \\ \hline
Building-3 & Office & 3 & 10 & floor \\ \hline
Building-4 & Residential & 4 & 20 & floor \\ \hline
\end{tabular}
\end{table*}

\begin{table*}[!tbp]
\centering
\caption{Details of steel foot overbridge where sensor module is deployed}
\label{steeloverbridgeinfo}
\begin{tabular}{|c|c|c|c|c|} \hline
Foot overbridge & Type of overbridge & Crowd density & Duration of data collection (minutes) & Position of deployment\\ \hline
Overbridge-1 & Steel-made & High & 10  & Middle position between 2 pillars \\ \hline
Overbridge-2 & Steel-made & Medium & 10 & Middle position between 2 pillars \\ \hline
Overbridge-3 & Steel-made & low  & 20 & Middle position between 2 pillars \\ \hline
Overbridge-4 & Steel-made & High & 10  & Middle position between 2 pillars \\ \hline
\end{tabular}
\end{table*}

\begin{table*}[!tbp]
\centering
\caption{Details of concrete foot overbridge where sensor module is deployed}
\label{concreteoverbridgeinfo}
\begin{tabular}{|c|c|c|c|c|} \hline
Foot overbridge & Type of overbridge & Crowd density & Duration of data collection (minutes) & Position of deployment\\ \hline
Overbridge-5 & Concrete-made & High & 10 & Middle position between 2 pillars \\ \hline
\end{tabular}
\end{table*}

\begin{table*}[!tbp]
\centering
\caption{Details of railline where sensor module is deployed}
\label{raillineinfo}
\begin{tabular}{|c|c|c|c|c|} \hline
Railline & Line type & \#Tracks & Duration of data collection (minutes) & Position of deployment\\ \hline
Railline-1 & Meter and broad gauge & 2 & 30 & attached with steel block \\ \hline
Railline-2 & Meter and broad gauge & 2 & 20 & attached with steel block \\ \hline
\end{tabular}
\end{table*}

\begin{table*}[!tbp]
\centering
\caption{A small portion of our dataset}
\label{sampledataset}
\resizebox{\textwidth}{!}{
\begin{tabular}{|c|c|c|c|c|c|c|c|c|c|c|c|c|} \hline
Mean & Mode & Median & \begin{tabular}{@{}c@{}}Standard \\ deviation\end{tabular}  & Max & Min & RMS & \begin{tabular}{@{}c@{}}Number of \\ peaks\end{tabular} & \begin{tabular}{@{}c@{}}Average of \\ peak values\end{tabular}  & Skewness & Kurtosis & \begin{tabular}{@{}c@{}}Creast \\ factor\end{tabular}  & \begin{tabular}{@{}c@{}}Type of \\ structure\end{tabular} \\ \hline
20.16 &	19 & 21 & 10.38 & 96 & 0 & 22.68 & 651 & 26.59 & 1.9 & 10.03 & 4.23 & Building \\ \hline

28.21 &	0 & 0 &	35.72 &	255 &	0 &	45.52 &	579 &	66 &	1.68 &	3.68 &	5.6 &	Flyover \\ \hline

62.86 & 5 &	63 &	1.89 &	66 &	37 &	62.89 &	498 &	63.77 &	-4.81 &	49.46 &	1.05 &	Railline \\ \hline

154.67 & 0 & 0 &	178.66 &	747 &	0 &	236.31 &	613 &	297.05 &	1.19 &	0.85 &	3.16 &	Steel overbridge  \\ \hline
24.9 & 68 &	0 & 24.45 & 259 & 0 & 34.9 & 630 & 44.88 & 1.78 & 7.29 & 7.42 & Concrete overbridge  \\ \hline
\end{tabular}
}
\end{table*}

\begin{figure*}
    \begin{subfigure}[t]{0.32\textwidth}
        \frame{\includegraphics[width=0.95\linewidth]{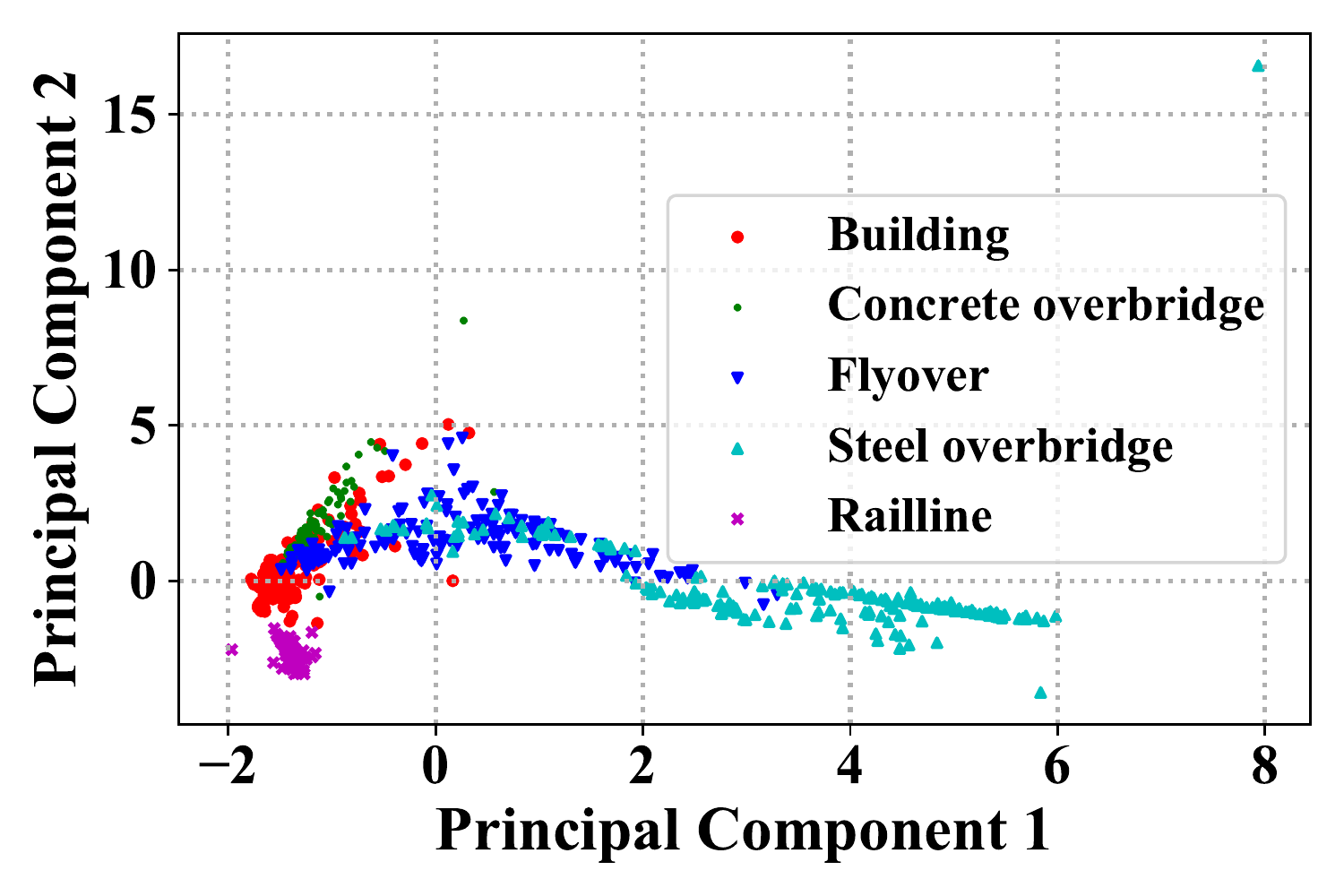}}
        \caption{Principle component analysis (PCA)}
        \label{pca}
    \end{subfigure}
    \begin{subfigure}[t]{0.32\textwidth}
       \frame{\includegraphics[width=0.95\linewidth]{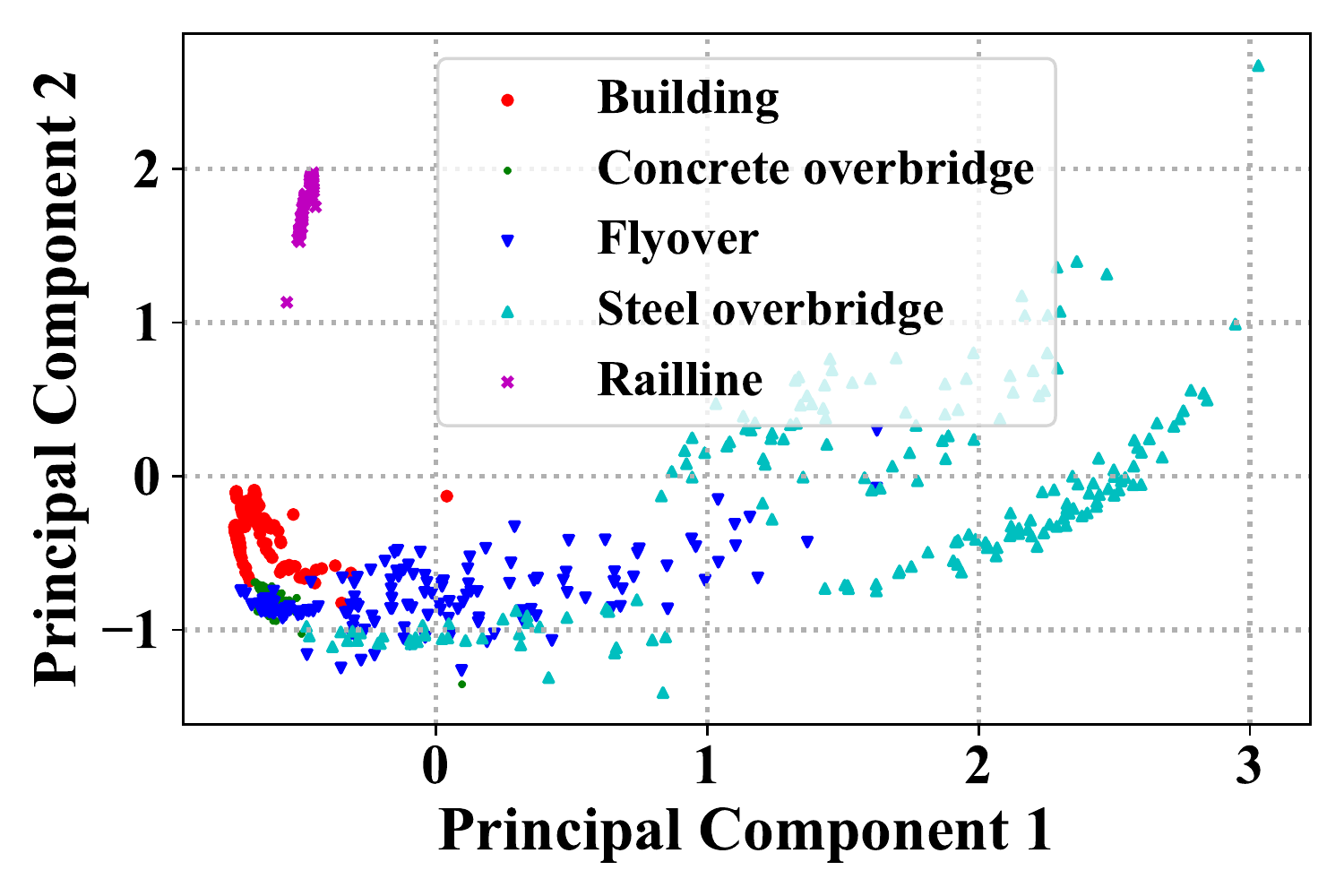}}
        \caption{Factor analysis}
        \label{fa}
    \end{subfigure}
    \begin{subfigure}[t]{0.32\textwidth}
        \frame{\includegraphics[width=0.95\linewidth]{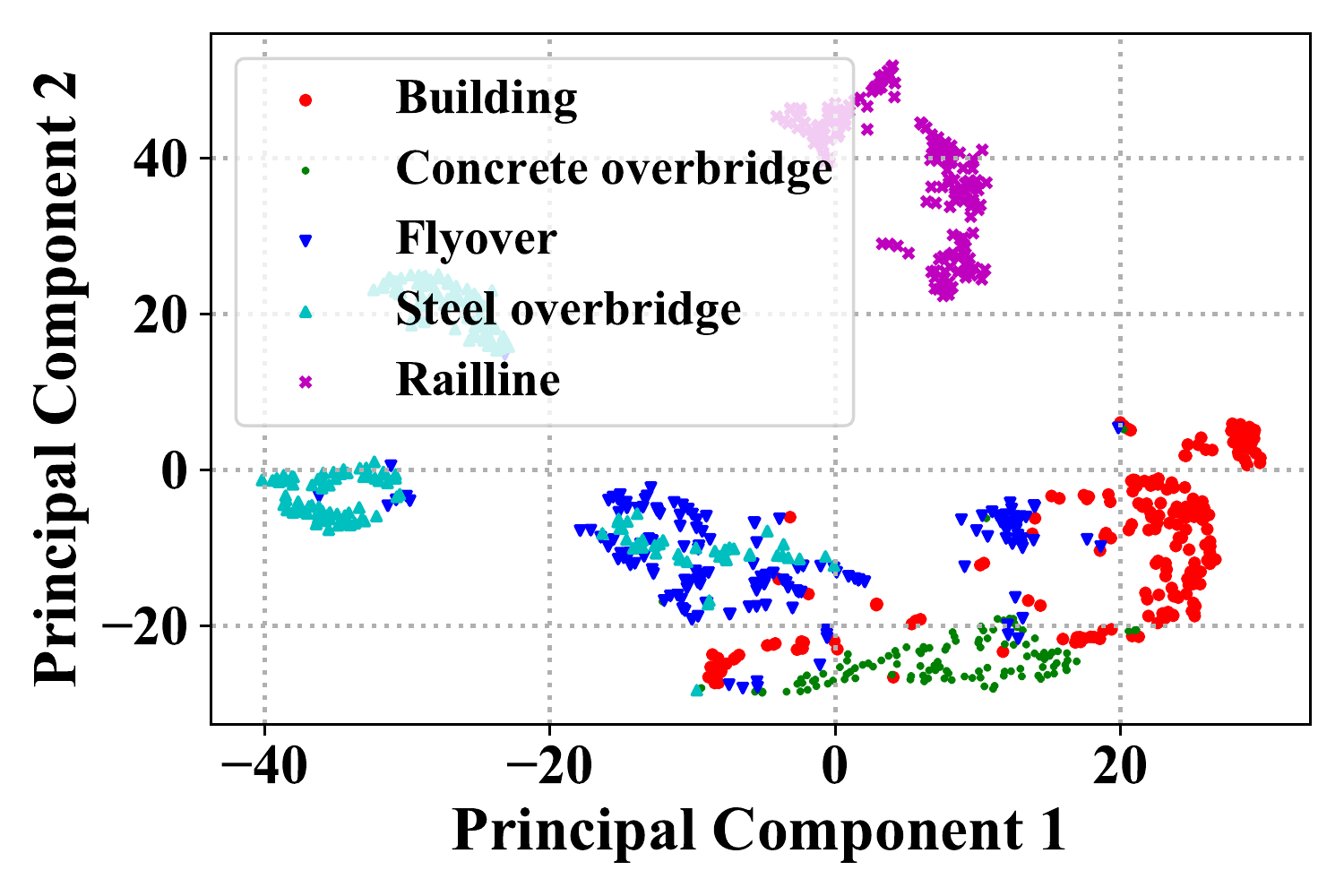}}
        \caption{T-distributed stochastic neighbor embedding (t-SNE)}
        \label{tsne}
    \end{subfigure}
    

\caption{Graphical representation of clusters formed by different structures}
\label{pca all}
\end{figure*}
\section{Data Visualization and Machine Learning Based Classification}
\label{dimensionality}
We perform different statistical and learning-based analysis over the collected data using scikit-learn version 0.23.1~\cite{scikit-learn}. The analyses help to visualize the data effectively and at the same time signifies the possibility of better classification.

To better visualize the data, we use PCA() method available in scikit-learn decomposition package ~\cite{pca}. This reduces feature dimension from 12 to two principal components and forms clusters of same type of structures as shown in Figure ~\ref{pca}. We also conduct T-distributed Stochastic Neighbor Embedding~\cite{tsne}, and Factor Analysis~\cite{fa} for better visualization. We present outcomes of all these analysis in Figure~\ref{pca all}. These figures clearly portrays that there is significant difference in vibration of the five structures.

\subsection{Classifying Structures from Vibration Data}

\begin{figure}[!tbp]
    \frame{\includegraphics[width=0.8\linewidth]{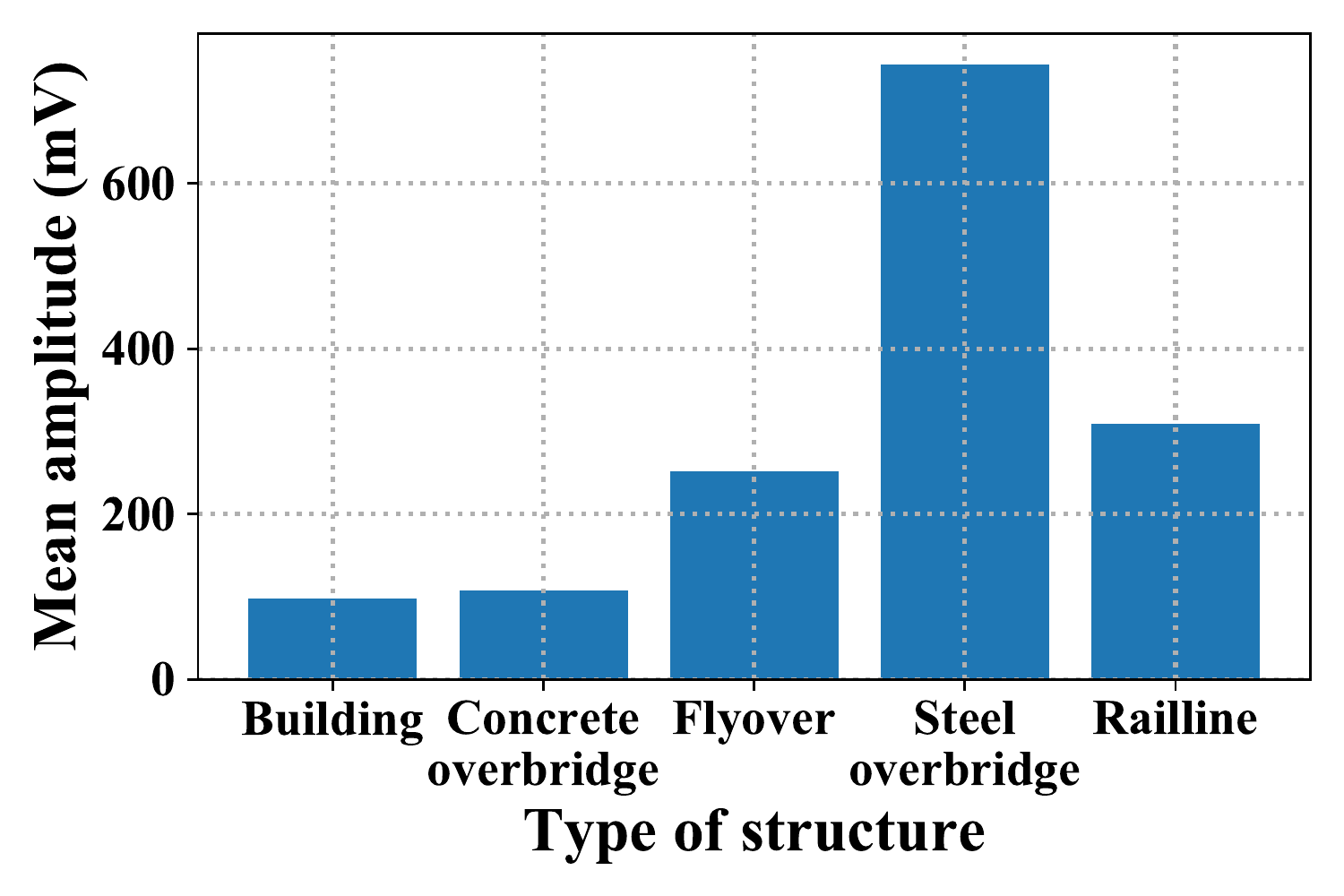}}
    \caption{Effect of structure type over mean \textit{amplitude} of vibration}
    \label{mean vs type}
\end{figure}

In this subsection, first, we show how the intensity of vibration varies with different structures. Then, we discuss how a feature value changes with the type of structures, i.e., which features are significantly affected by the type of structure.

\textbf{Vibration intensity versus type of structure: } Structure type  affects its generated vibration significantly, as materials used and design of architecture changes over the types of structures. As presented in Figure~\ref{mean vs type}, among the five different types of structures, concrete-made structures such as building and concrete overbridge are seemed to be less sensitive to generating vibration. On the other hand, steel-made structures such as steel overbridge and railline are much more sensitive to generating vibration. Though flyovers are made of concrete, due to heavy traffic flow, flyovers are also prone to generating substantial amount of vibration.

\textbf{Correlation between features and the type of structure:} We use Pearson's correlation coefficient (pearsonr() available in scipy stats package~\cite{pearson}) to identify- 1) how different statistical features and the type of structure are correlated with one another, and 2) whether there exist any statistically significant association (r >= 0.4 and p < 0.00005) between them or not. Here, we first generate the correlation matrix, and then then we determine prediction values of the regression matrix. In the correlation matrix, the feature having the highest absolute correlation coefficient value is highly related to the type of structure. On the other hand, in the regression matrix, feature with the least prediction value is highly significant to the type of structure.

\setlength{\parindent}{3ex} Table~\ref{correlation} shows the correlation and prediction values for all the features with the different types of structures. The table demonstrates that RMS, average of peaks, mean, standard deviation,  and max exhibit the highest correlation values (r >= 0.4). The same features also exhibit the lowest prediction values. Thus, we can deduce that RMS, average of peaks, mean, standard deviation, and max are highly significant features in terms of getting correlated. Accordingly, we conduct further analysis on classifying the type of structure using machine learning algorithms based on the selected five features.

\begin{table}[!tbp]
\centering
\caption{Correlation matrix and regression matrix(p-value) of type of structure with all features}
\label{correlation}
\begin{tabular}{|c|c|c|} \hline
Features & Correlation value & Prediction value\\ \hline
Mean &	0.716656053 &	     0.000000023\\ \hline
Median &	0.090901681 &	 0.190540928\\ \hline
Mode & 0.154416336	& 0.025589244\\ \hline
Standard deviation & 0.584570249 &	0.00000163\\ \hline
Max	& 0.406880713 &	         0.000000041\\ \hline
Min &	0.26979996 &	     0.000078\\ \hline
RMS &	0.784558941 &	     0.000000001\\ \hline
Number of peaks &	-0.274614018 &	0.0000572\\ \hline
Average of peaks &	0.756443394 &	0.000000017\\ \hline
Skewness &	0.228503295 &	 0.000875476\\ \hline
Kurtosis &	0.123029244 &	 0.075948388\\ \hline
Creast factor &	0.195535324 &	0.004549568\\ \hline

\end{tabular}
\end{table}

\textbf{Application of machine learning algorithms:} We apply several machine learning algorithms on our prepared dataset. Here, we formulate the task of predicting the type of structure from associated feature values as a classification problem where each class corresponds to one of the five different structures. The accuracy in our case corresponds to the number of correctly classified instances over the number of total test instances. We calculate different performance metrics such as precision, recall, and F-measure  in this regard. We use 10-fold cross-validation~\cite{crossval} for training each model. Then we conduct testing of each classifier model on unseen data points. In all cases, we maintain the ratio between training and testing dataset as 8:2.

\begin{figure}[!tbp]
    \frame{\includegraphics[width=0.8\linewidth]{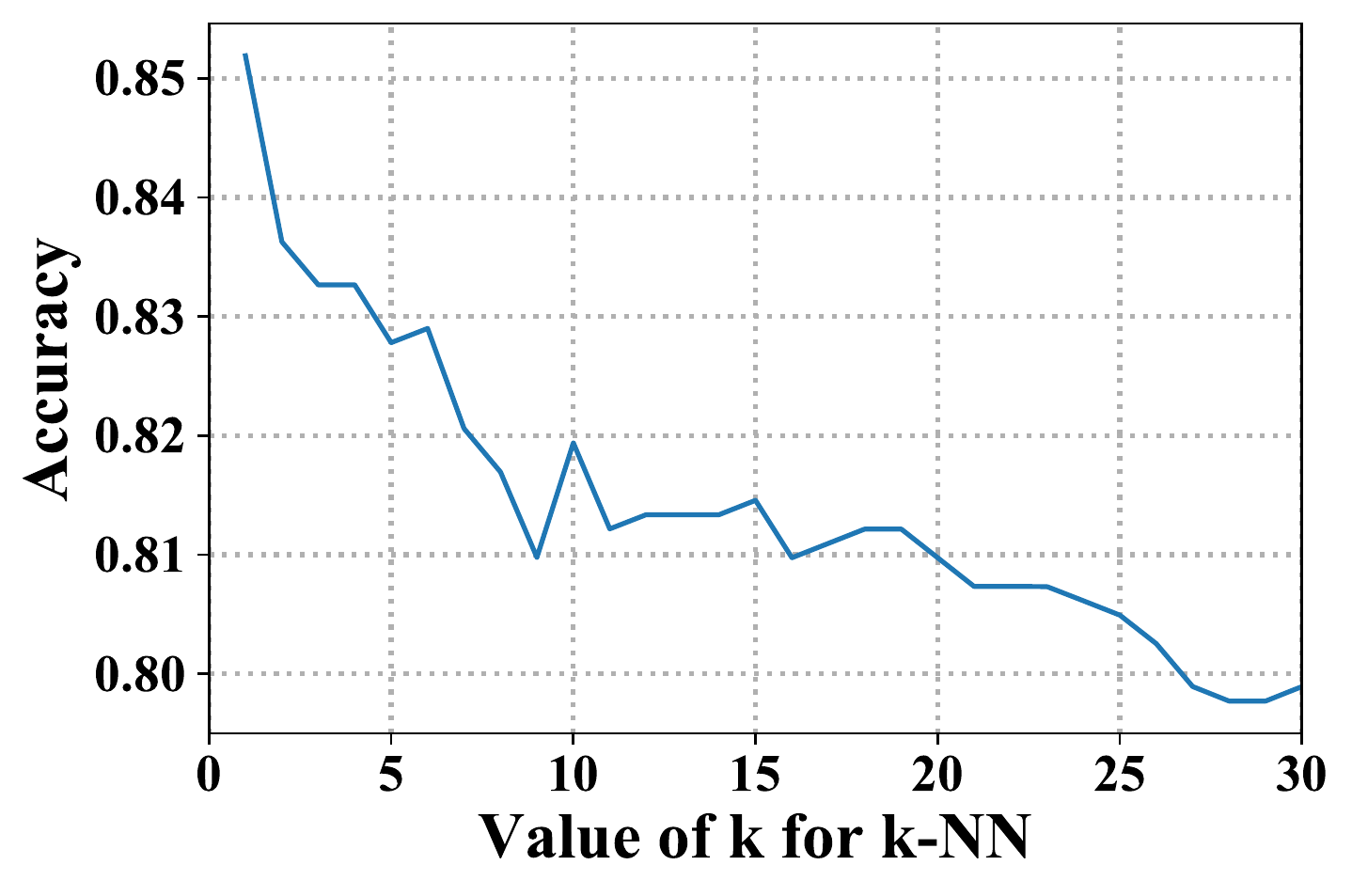}}
    \caption{Optimizing k for k-NN}
    \label{knn_opt}
\end{figure}

\setlength{\parindent}{3ex} Table ~\ref{performance} presents performance of all the classifiers under consideration. Among these classifiers, k-NN,  RandomForest, RandomTree perform the best in the metrics of  accuracy, precision, recall and F-measure. Among them, k-NN (for k=1) shows 91\% accuracy and outperforms others. For optimizing the value of k, here,  we cross-validate the k-NN model for k out the range from 1  to 30 with the training dataset. Our cross-validation results show that when the value for k is 1, validation accuracy exhibits the highest value as shown in Figure~\ref{knn_opt}.

\begin{figure}[!tbp]
    \frame{\includegraphics[width=0.8\linewidth]{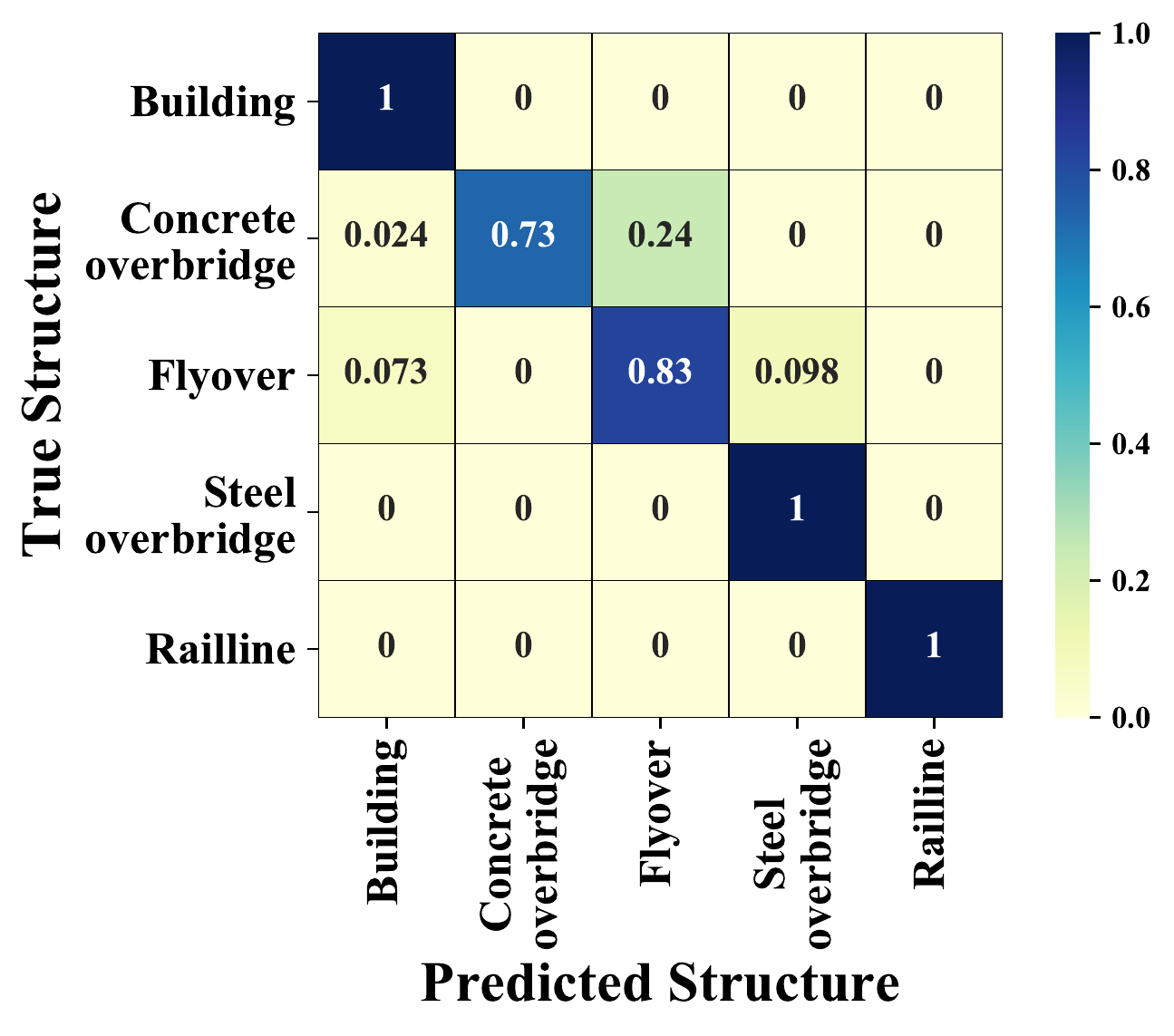}}
    \caption{Normalized confusion matrix}
    \label{confusion}
\end{figure}

\setlength{\parindent}{3ex} Figure~\ref{confusion} presents a normalized confusion matrix for k-NN (k=1). Here, among five structures, flyover gets misclassified as building or steel overbridge several times. Also, concrete overbridge gets misclassified as building and flyover in some cases. This leads to a high false-positive rate for flyover and concrete overbridge. The cause behind happening this is lower number of data points for flyover and concrete overbridge as we cover only one flyover and one concrete overbridge in our data collection phase. Besides, another reason is fact that both concrete overbridge and flyover are made of concrete. Thus, there can be a similarity of vibration for these two types of structures. Building, steel overbridge and rail line, on the other hand, get no false positive or false negative case. This is because we have substantial amount of data points for building, steel overbridge and railline. Also, vibration propagates more through metal structures, and, more importantly, in a more distinctive manner. Now, as there exists substantial room for further improvement, we employ Deep Learning for this purpose next.

\section{Deep Learning Based Approach}
\label{deep learning}
The major drawback of the aforementioned classical machine learning based approach is that it requires complex feature engineering. Among the 12 features in our dataset, we choose five features according to the correlation between the features and target classes. However, all models exhibit at most 90\% accuracy except k-NN. Even k-NN exhibits substantial error in classifying two classes (flyover and concrete overbridge). This suggests that a more rich and complex feature engineering might be required to achieve better performance. In this regard, Deep Neural Networks have emerged as a promising alternative for analyzing sensor data in recent times ~\cite{ravi2016deep, wang2019deep}. The main advantage of Deep Learning based approach is that it is completely data-driven and no manual feature engineering is required. In this regard, we propose a customized Deep Neural Network which is suitable for our desired task. In the following two subsections, first, we describe our model architecture, and then we present the experimental results.

\begin{table}[!tbp]
\centering
\caption{Performance matrix of some classifiers}
\label{performance}
\resizebox{\linewidth}{!}{
\begin{tabular}{|c|c|c|c|c|} \hline
Classifier & Accuracy(\%) & Precision & Recall & F-Measure\\ \hline
k-NN(k=1) & 91 & 0.92 & 0.91 & 0.91 \\ \hline
RamdomForest & 90 & 0.92 & 0.90 & 0.90 \\ \hline
RandomTree & 89 & 0.90 & 0.89 & 0.89 \\ \hline
Bagging & 84 & 0.88 & 0.84 & 0.84 \\ \hline
DecisionTable & 80 & 0.82 & 0.80 & 0.79 \\ \hline
NaiveBayes & 71 & 0.66 & 0.72 & 0.67 \\ \hline

\end{tabular}
}
\end{table}

\subsection{Model Architecture}

\begin{figure*}
\centering
\frame{\includegraphics[width=.9\linewidth]{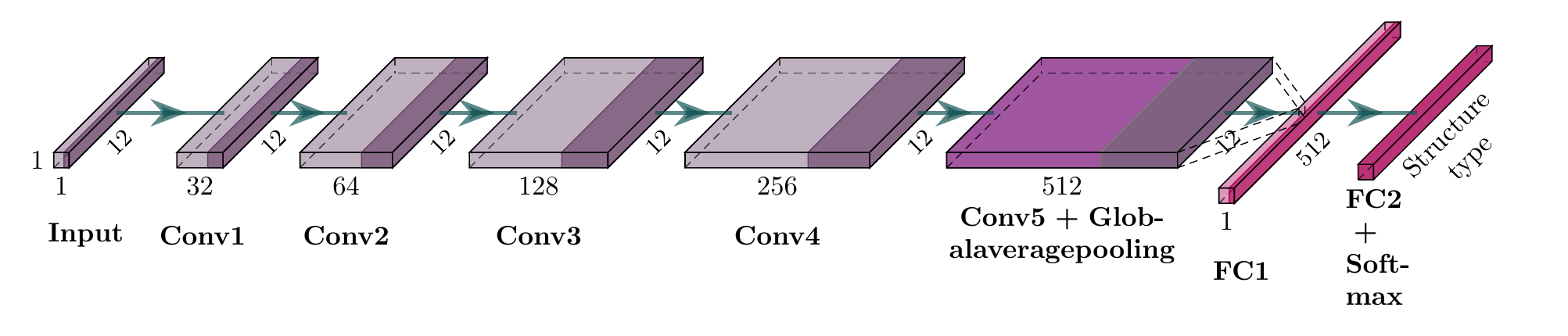}}
\caption{Architecture of our proposed Deep Learning based model}
\label{arch}
\end{figure*}

\begin{figure*}
\begin{minipage}[b]{0.24\textwidth}
  \begin{subfigure}{\textwidth}
    \centering
    \frame{\includegraphics[width=\textwidth]{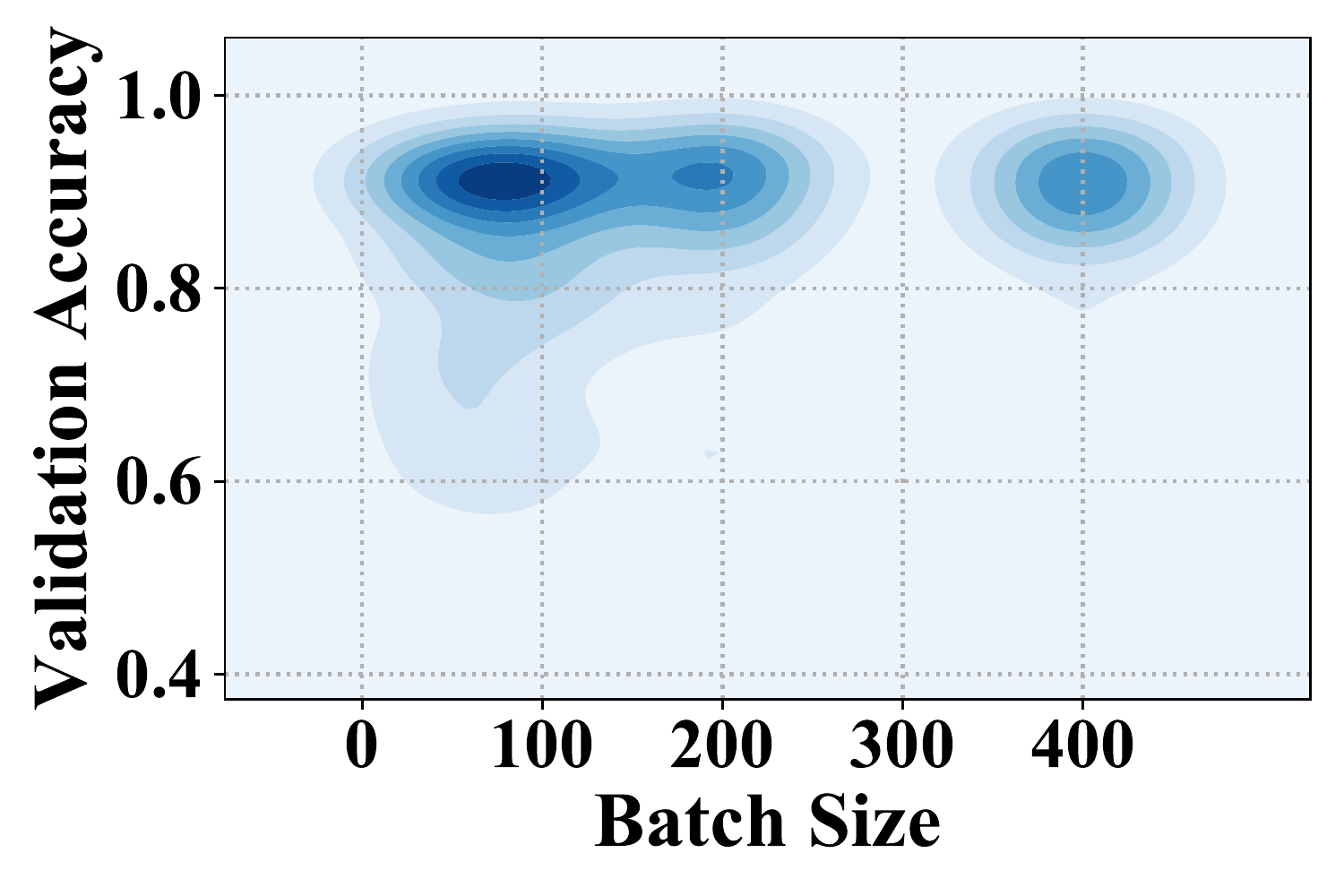}}
    \caption{Effect of batch size}
  \end{subfigure}
\end{minipage}
\hfill
\begin{minipage}[b]{0.24\textwidth}
  \begin{subfigure}{\textwidth}
    \centering
    \frame{\includegraphics[width=\textwidth]{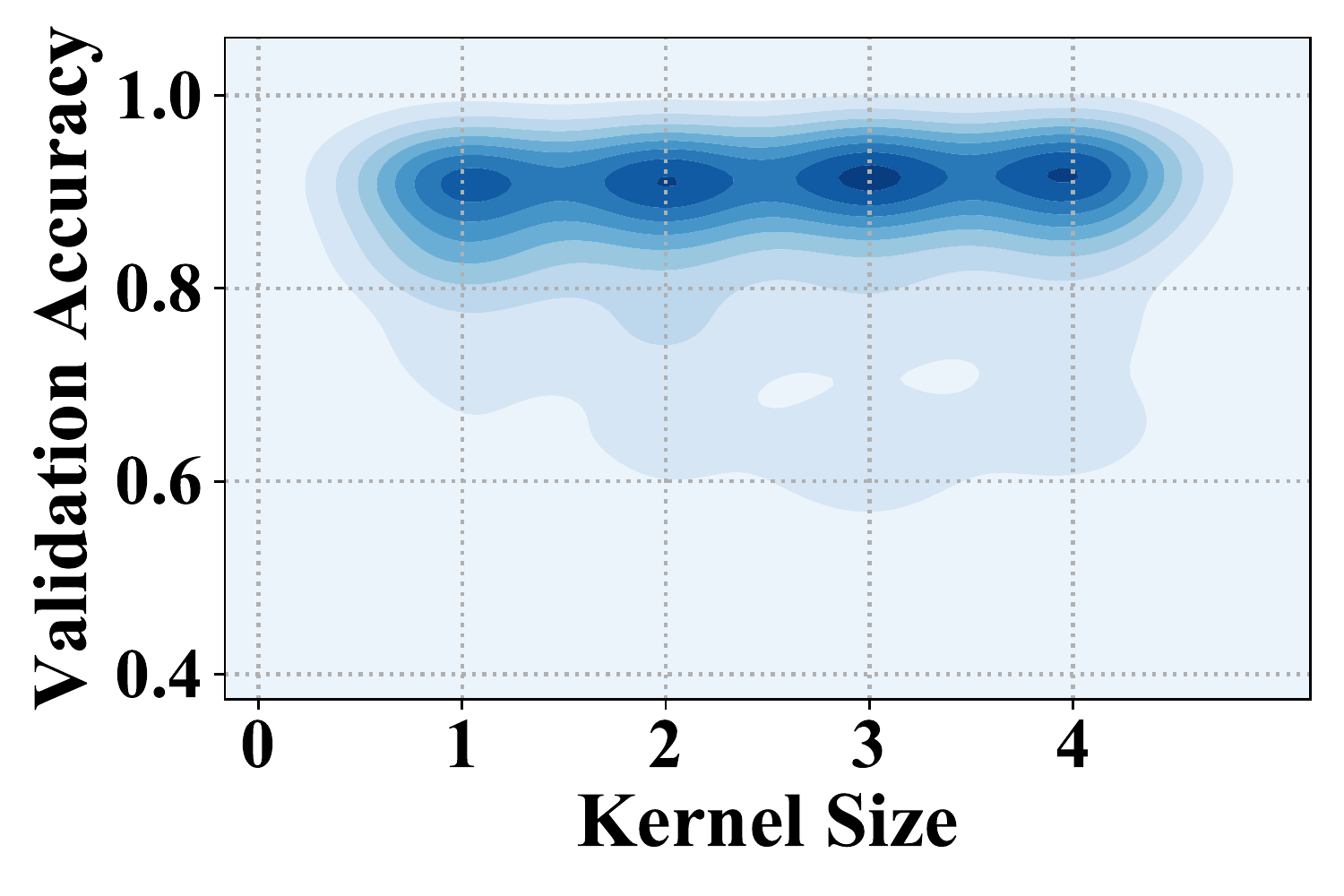}}
    \caption{Effect of kernel length}
  \end{subfigure}
\end{minipage} 
\hfill
\begin{minipage}[b]{0.24\textwidth}
  \begin{subfigure}{\textwidth}
    \centering
    \frame{\includegraphics[width=\textwidth]{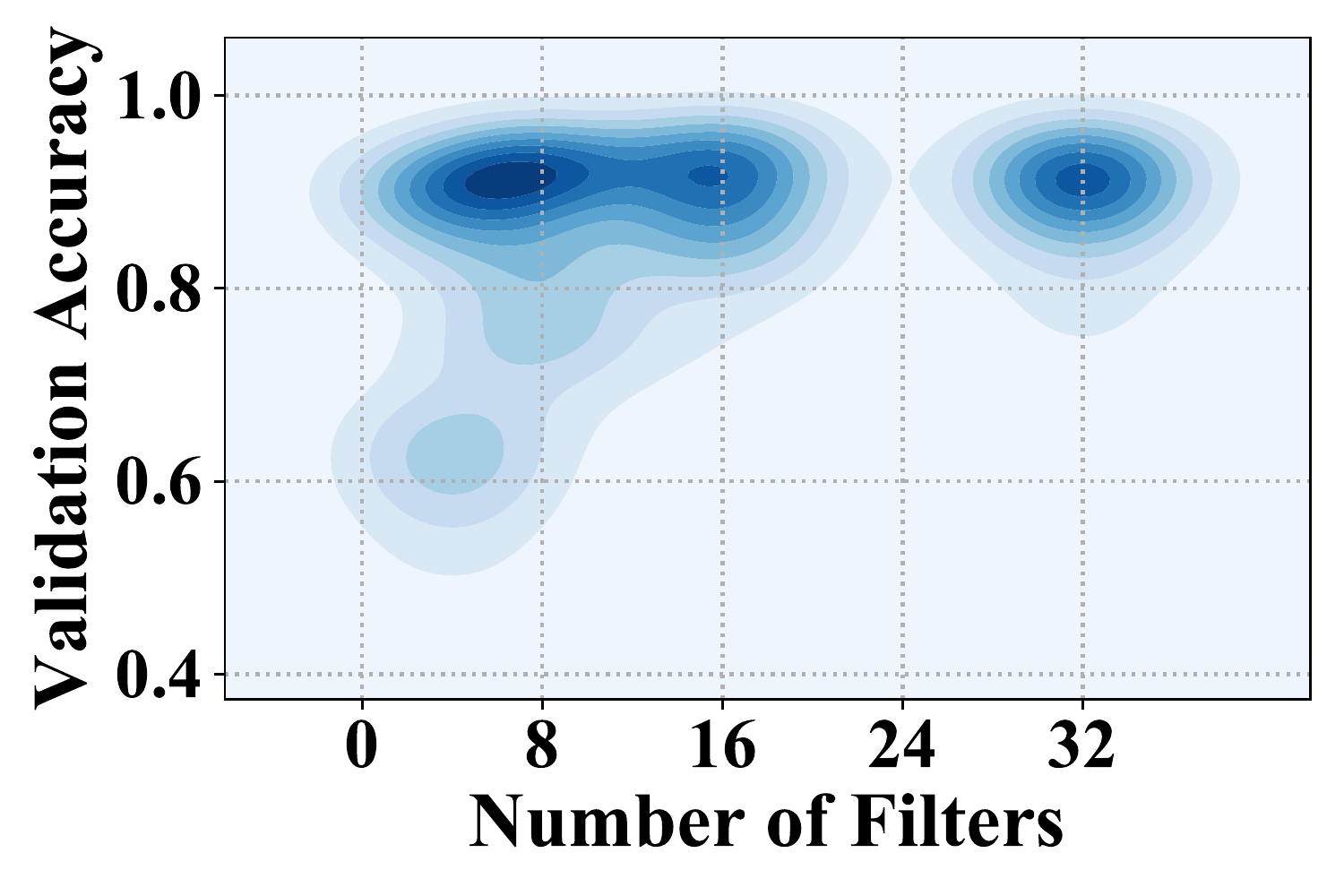}}
    \caption{Effect of filter numbers}
  \end{subfigure}
\end{minipage}
\hfill
\begin{minipage}[b]{0.24\textwidth}
  \begin{subfigure}{\textwidth}
    \centering
    \frame{\includegraphics[width=\textwidth]{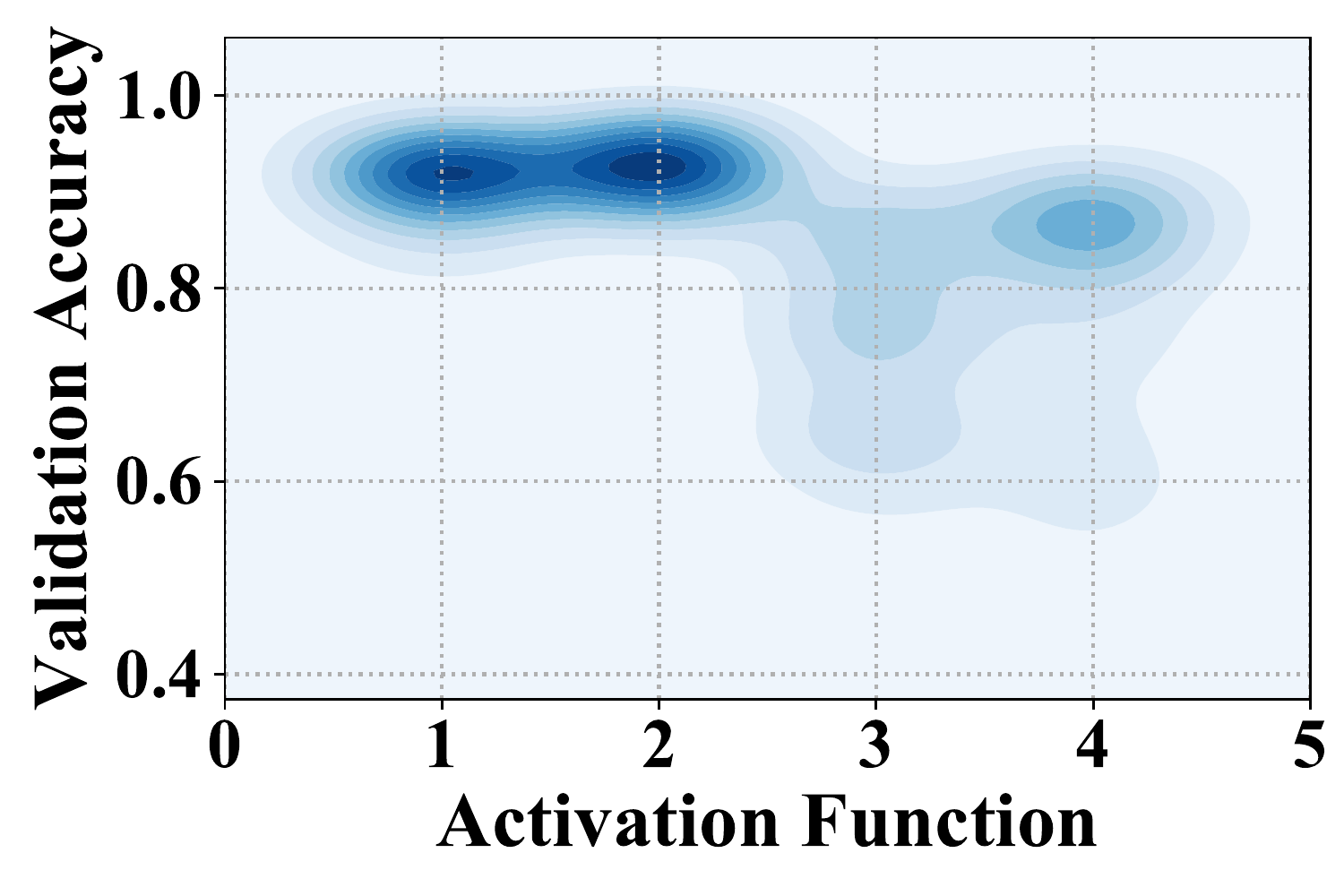}}
    \caption{Effect of activation function}
  \end{subfigure}
 \end{minipage} 
\caption{Kernel density estimation plot of the 256 experiments for each of the hyperparameters. In Figure (d), the number 1, 2, 3 and 4 indicate `ReLU', `ELU', `Tanh' and `Sigmoid' activation function respectively. }
\label{experiments}
\end{figure*}


Our proposed Deep Neural Network is based on Convolutional Neural Network (CNN). 
Even though CNN recently achieves a notable amount of success in computer vision and natural language processing~\cite{junior2018first, sindagi2018survey, al2018deep, khan2019survey}, research studies also show substantial success for sensor data analysis~\cite{fakhrulddin2017convolutional, essiet2019big, kamal2019intelligent}. The main advantage of CNN is that it can automatically extract necessary features from raw input data, which are typically hard to obtain through handcrafted feature engineering. A typical CNN consists of an input and an output layer, as well as multiple hidden layers. The hidden layers of a CNN typically consist of a series of convolutional layers that convolve with multiplication or other dot product to generate output feature maps. 



The convolution block of our model consists of five convolutional layers. The input shape to this convolution block is $n\times1\times12$ where \(n\) is the batch size and the number \(12\) is for all the 12 features from raw vibration data. The kernel size for each convolutional layer is $3\times3$. To learn a rich set of features, we increase the number of filters exponentially with the depth of the layers. The number of filters at the \(r^{th}\) convolutional layer is \(2^r\times q\) where \(0<=r<5\) and the value of \(q\) is selected as 32. The convolution operations are usually followed by activation functions that introduce non-linearity in the network. In our proposed model, we use the Exponential Linear Unit (ELU) as our activation function, which is defined as:

\begin{equation}
f ( x ) =   \begin{cases}
                x & x > 0 \\
                \alpha*(e ^ { x }-1) & x \leq 0 
            \end{cases}
\end{equation}

Our choice of hyperparameters is explained in the following subsection. We use a Globalaveragepooling layer~\cite{global_average_pooling} after the convolutional block to minimize the learnable parameters. Finally, we use a fully connected layer with $N$ number of output neurons along with softmax activation function to map the $N$ class scores to $N$ probability values $p= [p_1,p_2,....,p_N]$ for each class, which sums up to 1. 
We present an overview of the whole architecture in Figure~\ref{arch} and table~\ref{dnn_param}.

\begin{table}[!tbp]

\centering
\caption{Network parameters}
\label{dnn_param}
\resizebox{\linewidth}{!}{
\begin{tabular}{|c|c|c|} \hline
Layers. & Output Size & Kernels  \\ \hline
Input & $1\times12$  & -\\ \hline
Conv1D \& elu & $12\times32$  & $f = 32, K = 3, s = 1$\\ \hline
Conv1D \&  elu & $12\times64$  & $f = 64, K = 3, s = 1$\\ \hline
Conv1D \&  elu & $12\times128$  & $f = 128, K = 3, s = 1$\\ \hline
Conv1D \&  elu & $12\times256$  & $f = 256, K = 3, s = 1$\\ \hline
Conv1D \&  elu & $12\times512$  & $f = 512, K = 3, s = 1$\\ \hline
Globalaveragepooling1D & $1\times512$  & -\\ \hline
Fully connected & $1\times N$  & -\\ \hline
\end{tabular}
}
\begin{tablenotes}
  \small
  \item * here $f$, $K$, $s$, and $N$ represent number of filters, kernel length, filter stride and number of classes respectively.
\end{tablenotes}

\end{table}

\subsection{Experimental Setup and Hyperparameter Tuning in Our Deep Learning}

The model hyperparameters for our network contains batch size, kernel size, number of filters, and activation function. Here, we vary the batch size as 50, 100, 200, and 400. Besides, we vary kernel length as 1, 2, 3, and 4. We also vary the number of filters ($q$) for the first convolution layer as 4, 8, 16, and 32. 

At first, we split the dataset into 70\% training set, 10\% validation set, and 20\% test set. We use 10-fold cross-validation to find the value the hyperparameters. This results in a total $10 \times 256 $ experiments. Here 10 is the total number folds. The number 256 is the number of possible combination of hyperparameters. Some possible combination of hyperparameters (batch size, kernel size, number of filters, activation function) are (50, 1, 4, Tanh), (50, 2, 32, ELU), (200, 2, 4, ReLU), etc. The average results of the 256 experiments over the 10 folds are shown as kernel density estimation plots in Figure~\ref{experiments}. It is evident that a combination of batch size 100, kernel length 3, number of filters 32, and `ELU' activation function achieves the highest validation accuracy. Based on these results, we set the hyperparameters in our model as (100, 3, 32, ELU). All of the experiments regarding training, testing, and hyperparameter tuning of the networks are performed in Kaggle kernel environments which provides Nvidia K80 GPUs~\cite{kaggle}. We write necessary codes in Python and implement the neural network models using the Keras API with TensorFlow in the back-end~\cite{keras, tensorflow}.

\subsection{Experimental Results and Analysis}

\begin{table}[!tbp]
\centering
\caption{Model performance on training phase}
\label{training_dnn}
\begin{tabular}{|c|c|} \hline
Training accuracy & Validation accuracy \\ \hline
97.7\%  &    96.7\%\\ \hline
\end{tabular}
\end{table}
\begin{table}[!tbp]
\centering

\caption{Model performance on testing phase}
\label{performance_dnn}
\begin{tabular}{|c|c|c|c|c|} \hline
Testing accuracy & Precision & Recall & F-Measure\\ \hline
97.1\% & 0.97 & 0.97 & 0.97 \\ \hline
\end{tabular}
\end{table}

\begin{figure}[!tbp]
    \frame{\includegraphics[width=.8\linewidth]{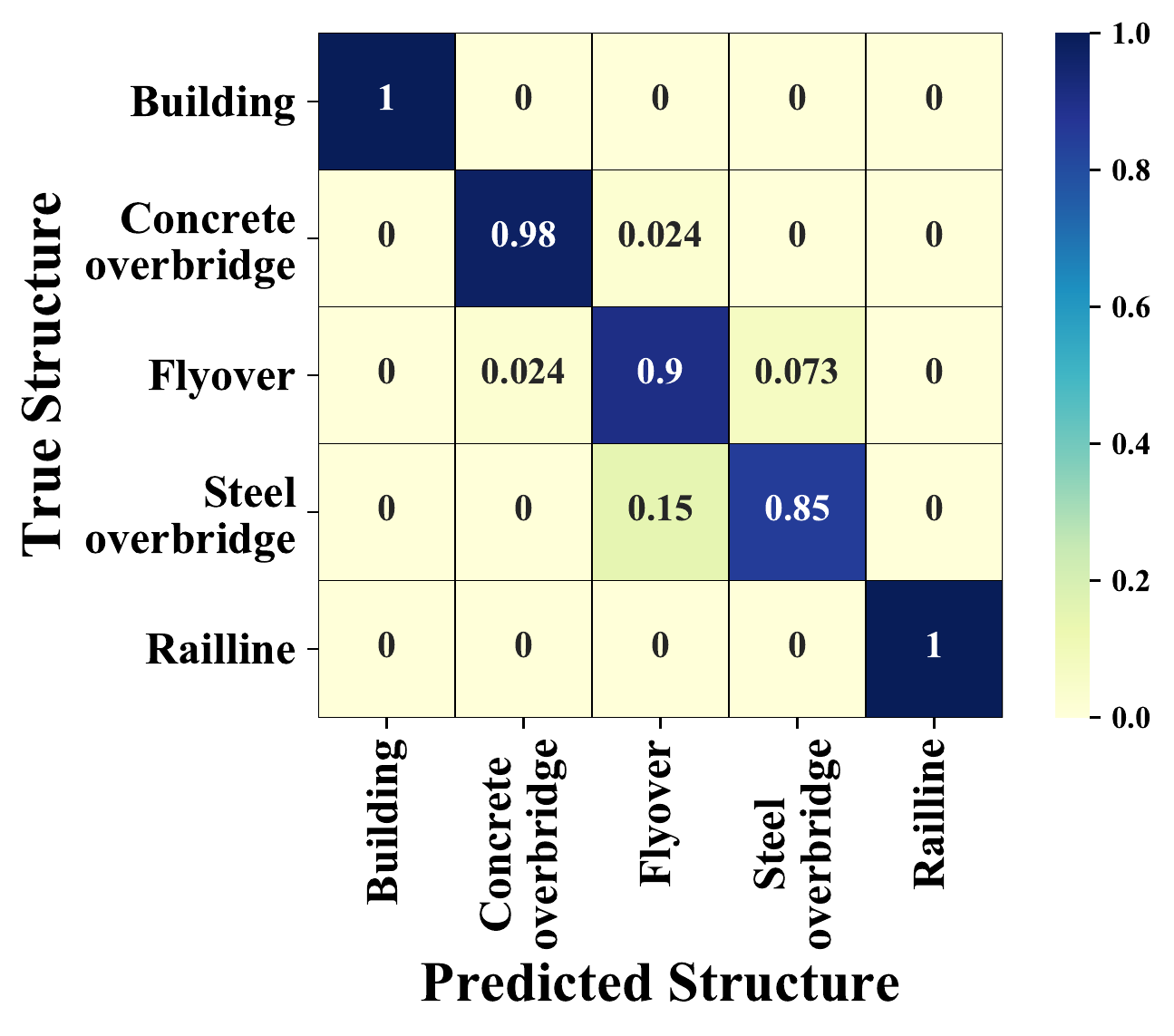}}
    \caption{Normalized confusion matrix}
    \label{confusion_dnn}
\end{figure}

We evaluate the performance of our Deep Learning based model over the collected dataset in two stages. At first, we evaluate the performance of our model in the training phase with 10-fold cross-validation. In each fold, we train the model for 1000 epochs. We use Adam~\cite{adam} as an optimizer with an initial learning rate of \(10^{-2}\). We also use a learning rate decay factor of 0.8 if the validation accuracy does not improve for 10 consecutive epochs~\cite{epoch}. Table~\ref{training_dnn} represents the average training accuracy and validation accuracy of this experiment.

\setlength{\parindent}{3ex} In the second stage, we evaluate our model over unseen test data, which can be of untrained building, railline, steel overbridge, and concrete overbridge. As we collect data from only one flyover, we use it for both training and testing. Among the 10 models from every 10 folds, we choose the best model having the highest validation accuracy. Then, we evaluate the model over several performance metrics, such as accuracy, precision, recall, and F1-score. Our Deep Learning based model outperforms the best found machine learning based k-NN (91\%) in terms of all performance metrics. Table~\ref{performance_dnn} presents values of all performance metrices.

\setlength{\parindent}{3ex} Figure~\ref{confusion_dnn} presents a normalized confusion matrix for the test set evaluation. Among the five structures, flyover gets misclassified as concrete and steel overbridge for few times, though the false positive rate here is less compared to machine learning based approach as in Figure~\ref{confusion}. This happens as we cover only one flyover in our data collection phase, resulting in a relatively smaller amount of data. Besides, both concrete overbridge and flyover are made of concrete, and thus there can be a similarity of vibration from these two structures. Nonetheless, building and railline get no false positive or false negative case.

\setlength{\parindent}{3ex} The reason for Deep Learning based approach performing better is that Deep Learning does not require any feature selection procedure. On the other hand, in our machine learning based approach, we select five statistical features among 12 from our dataset according to our analysis on correlation and significance. However, in our Deep Learning based approach we take all of the 12 features ignoring their correlation and significance. This helps in learning of our model significantly, and thus, in achieving a higher accuracy.


\section{Variation in Vibration at Different Heights of A Building}
\label{section: vibration and building height}

\begin{figure}[!tbp]
\centering
    \frame{\includegraphics[width=.5\linewidth]{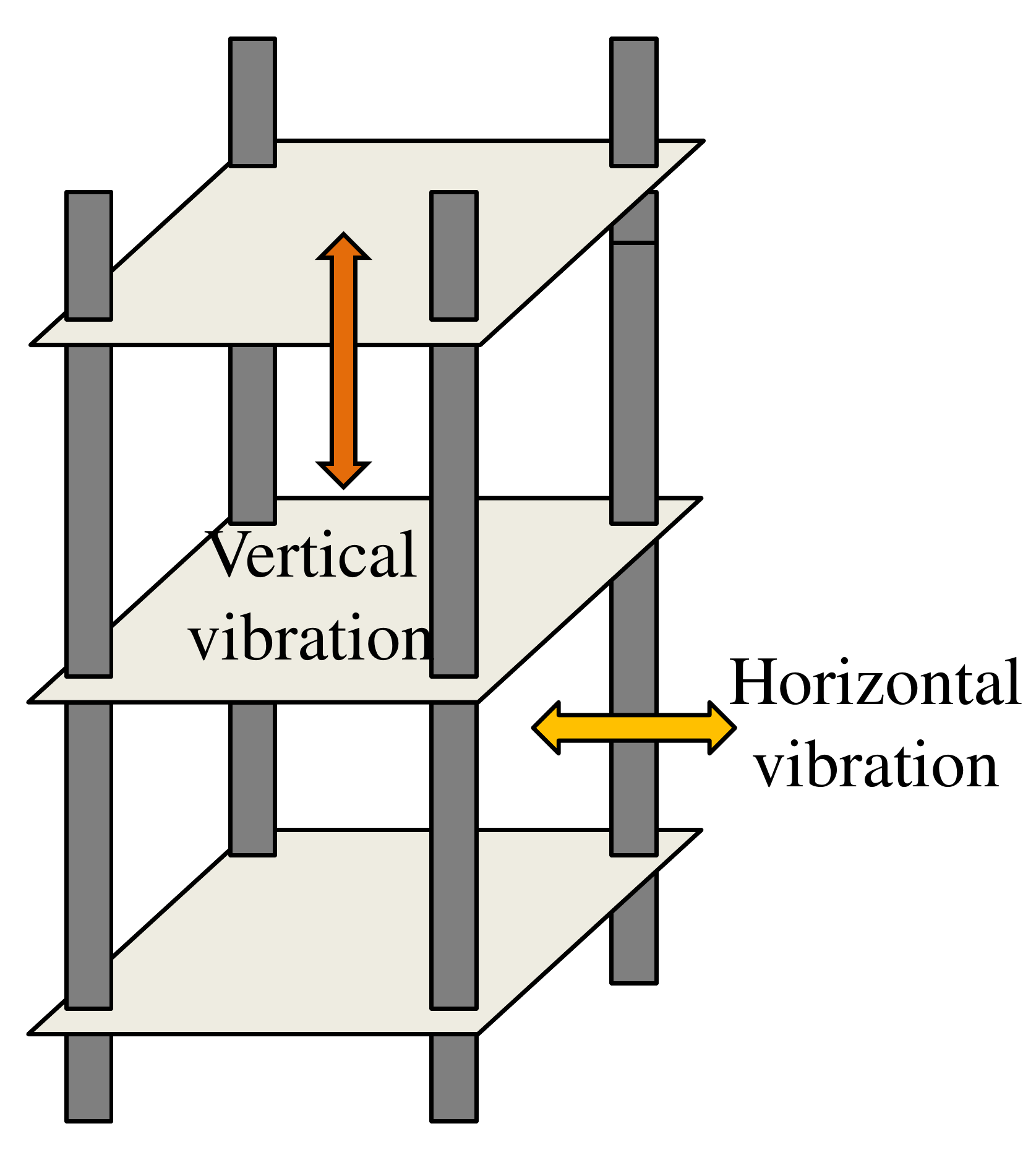}}
\caption{Vertical and horizontal vibration inside a building}
\label{horizontal and vertical deploy}
\end{figure}

\begin{figure}[!tbp]
    \centering
    \begin{subfigure}{0.23\textwidth}
        \frame{\includegraphics[width=\textwidth]{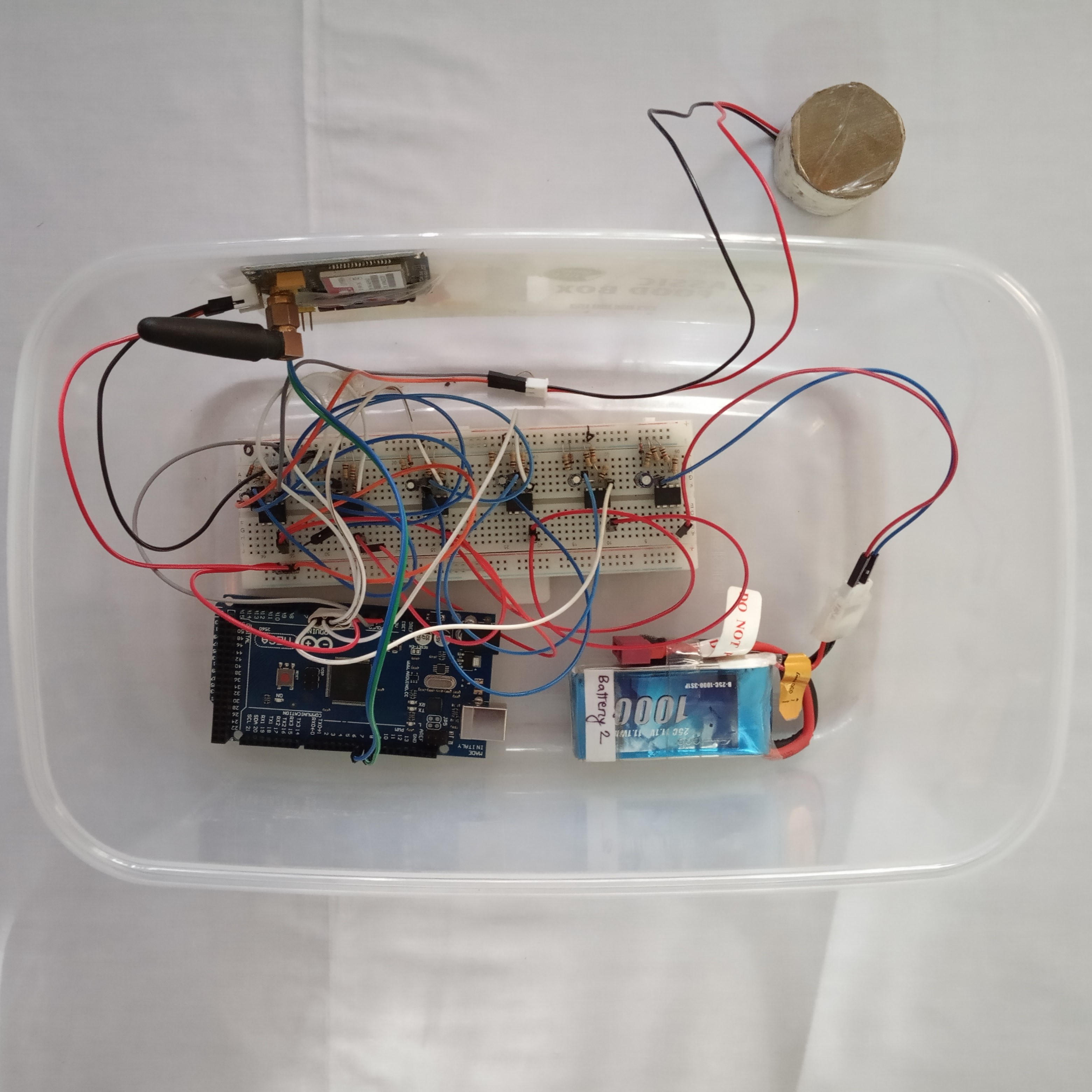}}
        \caption{Sensor on floor surface}
        \label{fig:vertical}
    \end{subfigure}%
    \hfill
    \begin{subfigure}{0.23\textwidth}
        \frame{\includegraphics[width=\textwidth]{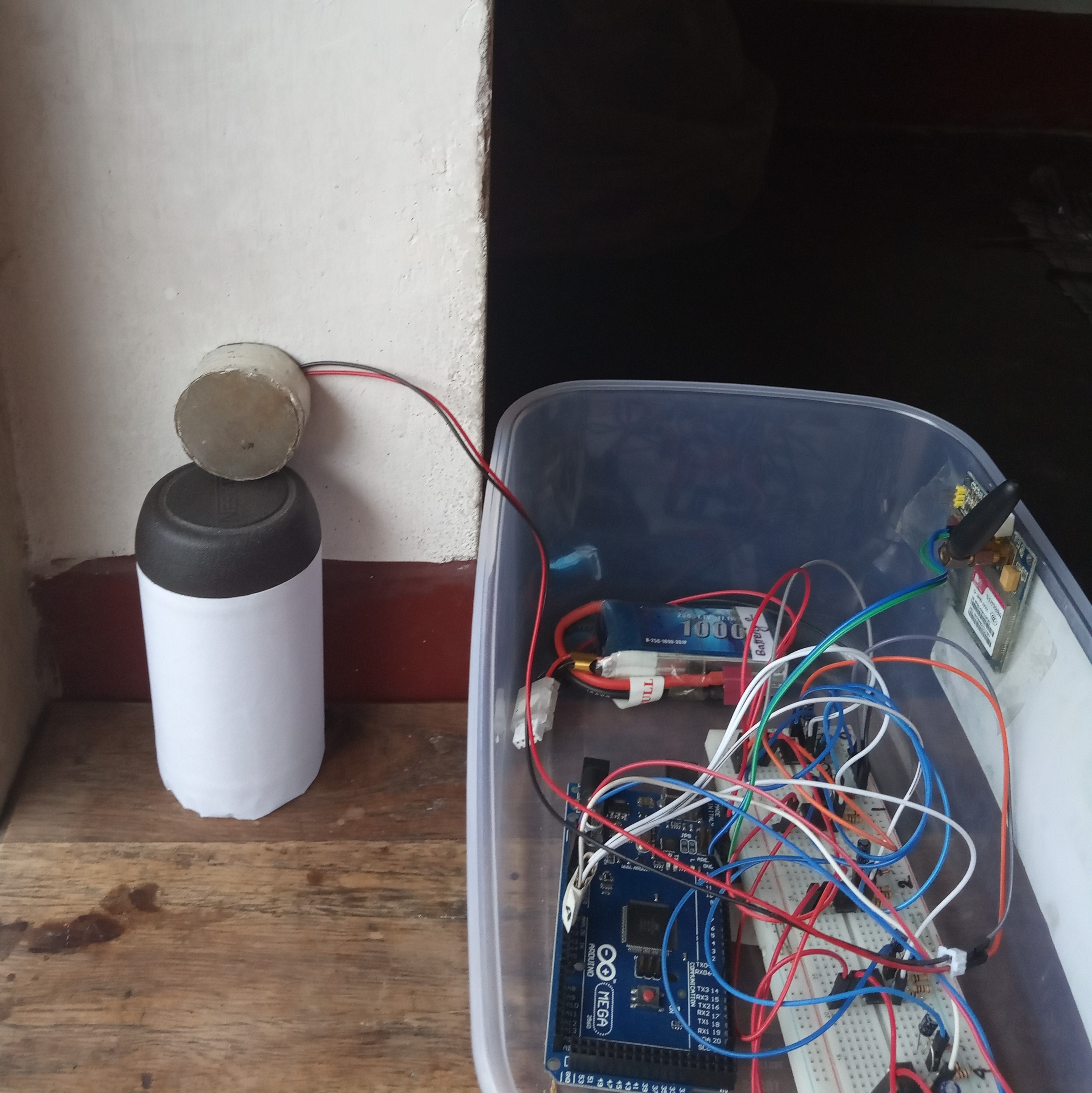}}
        \caption{Sensor on pillar surface}
        \label{fig:horizontal}
    \end{subfigure}%
        \caption{Deployment of sensing module inside a building}
        \label{fig:deployment}
\end{figure}

In addition to classifying structures based on their generated vibration, we dig into further on characteristics of vibration generated by a structure at its different heights. As a case study, we perform this analysis on buildings.

In buildings, vibration can be of two types - vertical and horizontal~\cite{bnbc1, bnbc2}. Figure~\ref{horizontal and vertical deploy} presents possible two types of vibrations for a dummy 2 storey building. Here, the vertical vibration acts on floors of a building and horizontal vibration acts on pillars (or columns) of the building. Foot strikes due to human walking and similarly varying vertical pressures cause the vertical vibration. On the other hand, wind pressure and similar varying horizontal pressures cause horizontal vibration. We attempt to study the relationship between the types of vibrations and the heights from where they get generated. Next, we describe detail methodology and experimentation of our study.

\subsection{Methodology of Our Study} To study the effect of floor height on vibration generated by the floor of a building, we consider two multi-storey buildings having the number of floors 11 and five in Dhaka city. We collect vertical and horizontal vibration data for 300 seconds in each floor of the two buildings. Then, we calculate the mean \textit{amplitude} of vibration and apply linear curve fitting to get a linear equation that best relates mean \textit{amplitude} of the vibration and the associated floor index. 

\subsection{Experimental Setup and Deployment}
For sensing vertical and horizontal vibrations of the buildings, we use the same hardware setup previously described in section~\ref{setup1}. However, here, we consider two different deployment scenarios - one for sensing vertical vibration and another for sensing horizontal vibration. For sensing vertical vibration, we deploy our sensing module on the surface of each floor as shown in Figure~\ref{fig:vertical}. For sensing horizontal vibration, we deploy our sensing module on the surface of a pillar as shown in Figure~\ref{fig:horizontal}. 
%

\subsection{Vertical Floor Vibration} 

\begin{figure}[!tbp]
\centering
    \begin{subfigure}[t]{0.23\textwidth}
    \centering
    \frame{\includegraphics[width=\linewidth]{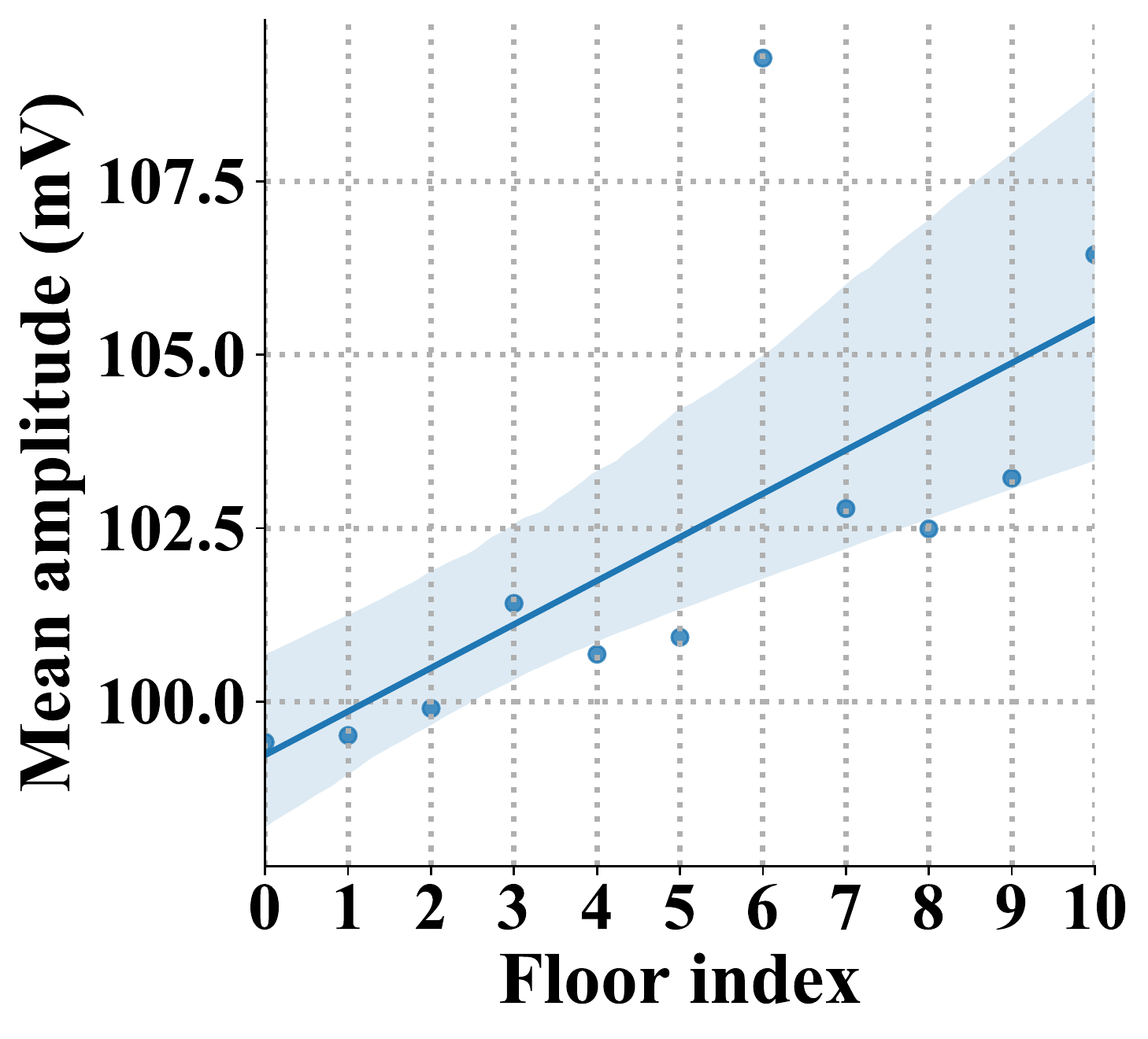}}
    \caption{Building-1 (11 storey)}
    \label{b1}
    \end{subfigure}
    \hfill
    \begin{subfigure}[t]{0.215\textwidth}
    \centering
    \frame{\includegraphics[width=\linewidth]{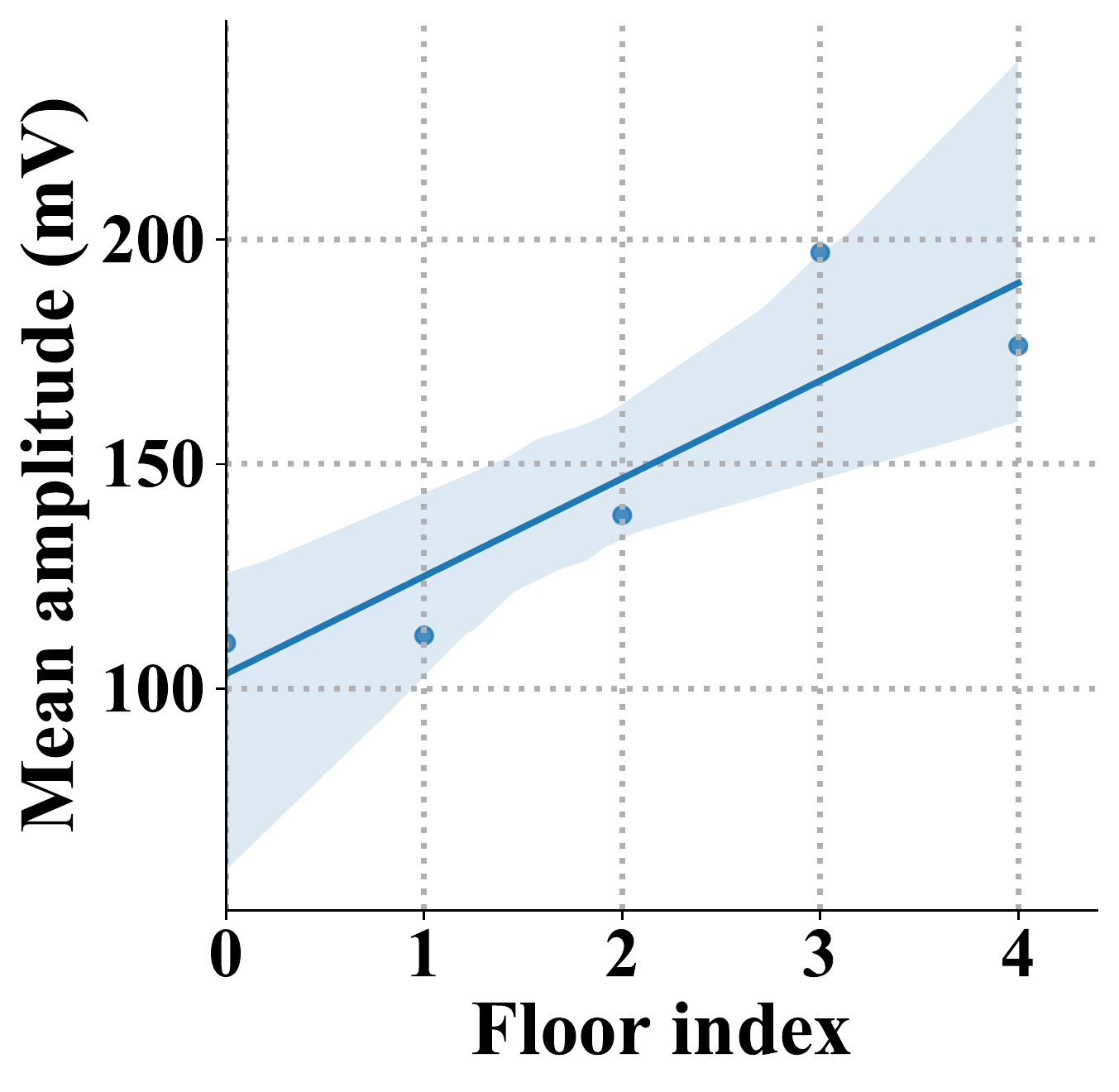}}
    \caption{Building-2 (5 storey)}
    \label{b2}
    \end{subfigure}
\caption{Vertical floor vibration versus floor index}
\label{mean vs floor}
\end{figure}

For the vertical deployment scenario as shown in Figure~\ref{fig:vertical}, we find two linear equations after applying linear curve fitting for the 11 and 5 storey buildings. For linear curve fitting, we use polyfit() function available in numpy~\cite{polyfit}. Figure~\ref{mean vs floor} shows the fitted lines, which confirms that both the derived equations are in $y = m \times x + c$ format.

\begin{equation}
\label{eq1}
    Mean\;\textit{amplitude} = 0.12 \times Floor\;index + 20.3
\end{equation}
\begin{equation}
\label{eq2}
    Mean\;\textit{amplitude} = 4.46 \times Floor\;index + 21.2
\end{equation}

\subsection{Horizontal Pillar Vibration}

For the horizontal deployment scenario as shown in Figure~\ref{fig:horizontal}, we again find two linear equations after applying our linear curve fitting for both the buildings. Figure~\ref{mean vs floor 2} shows the fitted lines that follow the equation format $y = m \times x + c$. Actual derived equations are as follows.
\begin{equation}
\label{eq3}
    Mean\;\textit{amplitude} = -0.2 \times Floor\;index + 28.2
\end{equation}
\begin{equation}
\label{eq4}
    Mean\;\textit{amplitude} = -0.6 \times Floor\;index + 29.9
\end{equation}

\subsection{Result Analysis}
\begin{figure}[!tbp]
    \begin{subfigure}[t]{0.23\textwidth}
    \frame{\includegraphics[width=\linewidth]{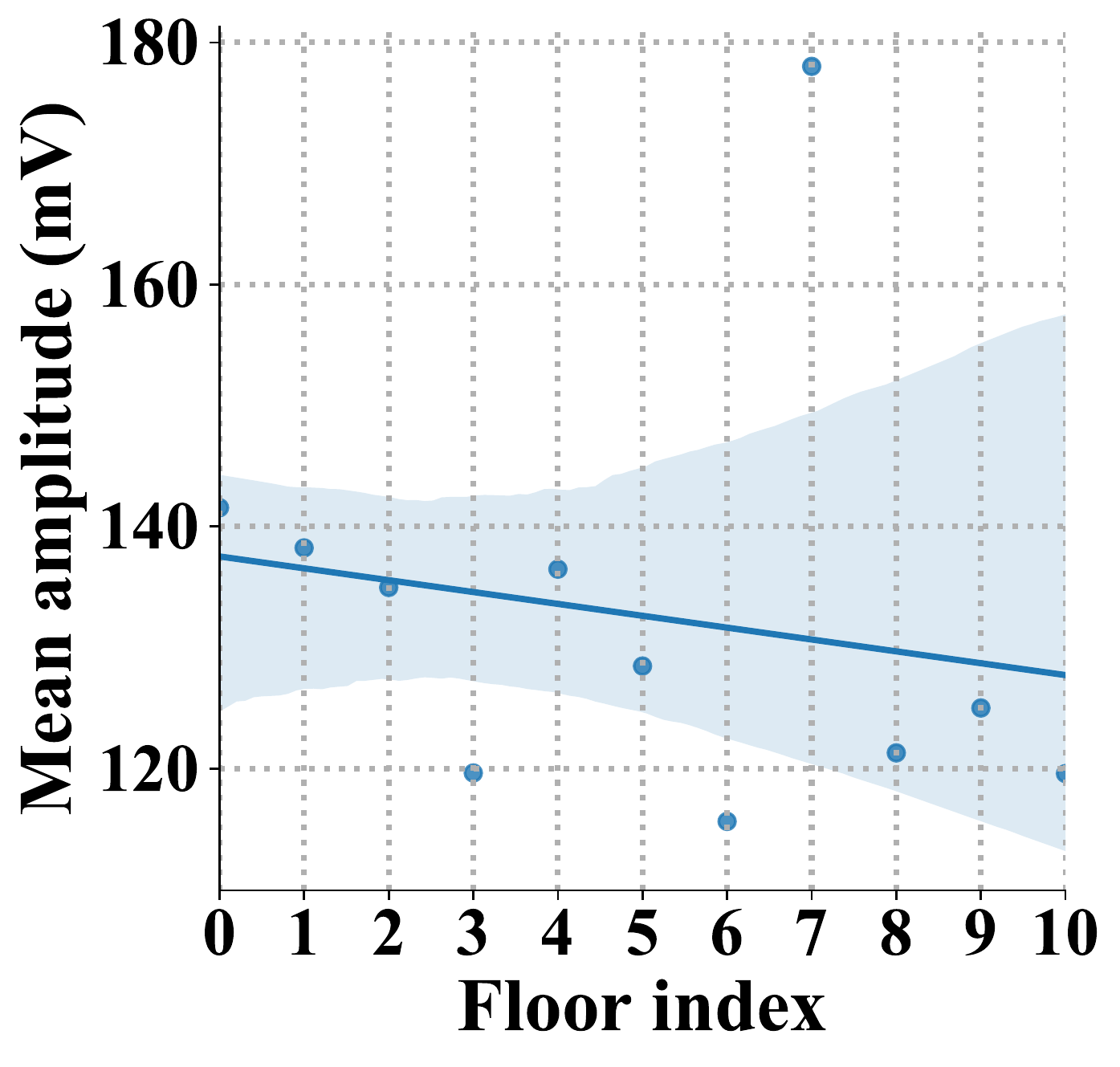}}
    \caption{Building-1 (11 storey)}
    \label{b1_}
    \end{subfigure}
    \hfill
    \begin{subfigure}[t]{0.225\textwidth}
    \frame{\includegraphics[width=\linewidth]{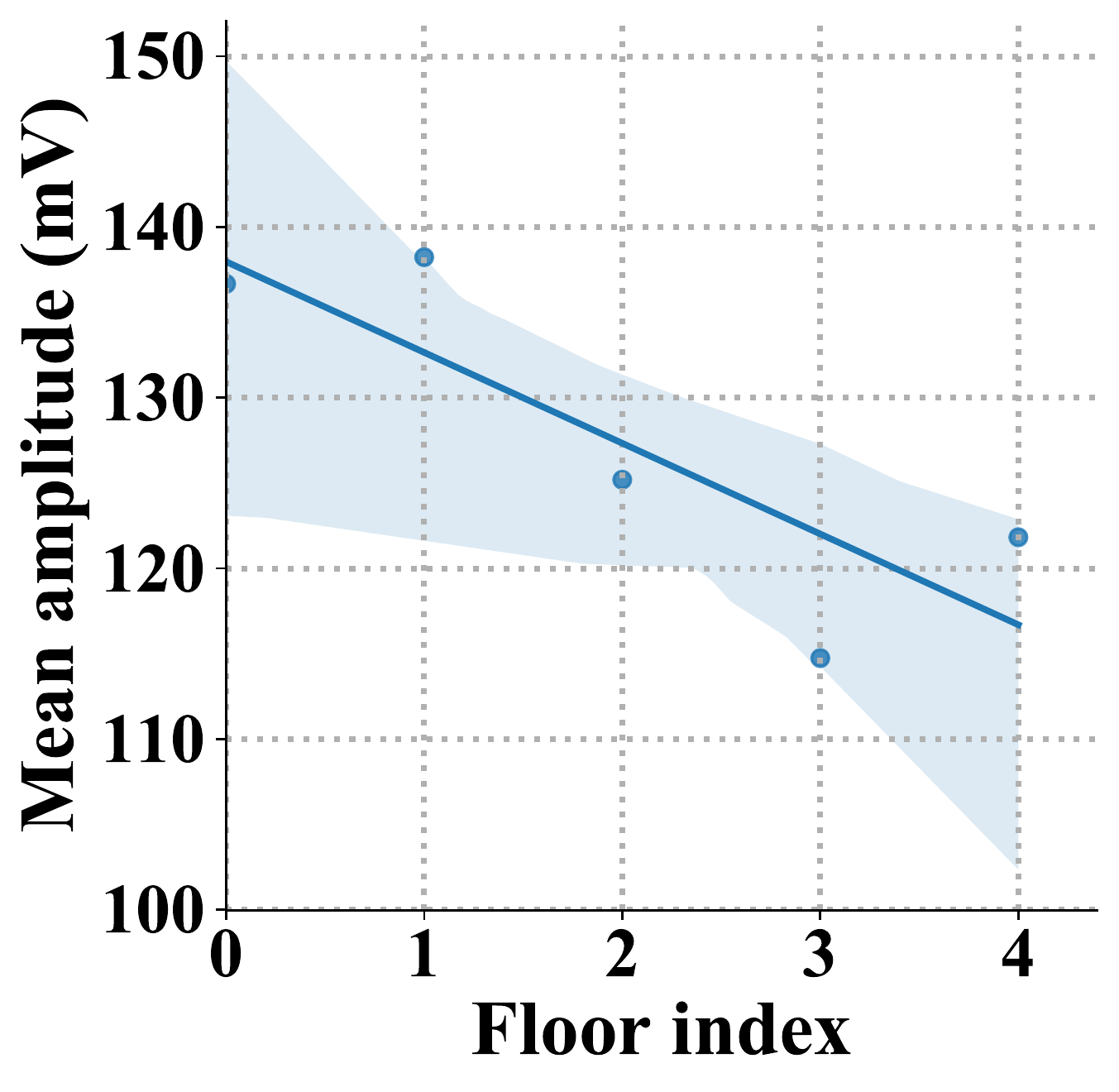}}
    \caption{Building-2 (5 storey)}
    \label{b2_}
    \end{subfigure}
\caption{Horizontal pillar vibration versus floor index}
\label{mean vs floor 2}
\end{figure}

Figure~\ref{mean vs floor} demonstrates that, as the floor height increases, mean \textit{amplitude} of vertical vibration also increases. Also, equation~\ref{eq1},\ref{eq2} quantify the extent of increase exhibiting positive slopes in both the cases. On the other hand, Figure~\ref{mean vs floor 2} demonstrates that, as the floor height increases, mean \textit{amplitude} of horizontal vibration decreases. Equation~\ref{eq3}, \ref{eq4} quantify the extent of decrease exhibiting negative slopes in both the case. To the best of our knowledge, we are the first to reveal such a finding on decaying \textit{amplitude} of horizontal vibration of pillars with an increase in the height at which the vibration is getting generated.

\section{Conclusion and Future Work}
Analyzing the dynamics of vibration for diversified civil structures is little explored in literature - specially from the perspective of using low-cost vibration sensing. Therefore, in this study, we analyze the dynamics in depth by devising and utilizing a low-cost vibration sensing module. Our sensing module continuously uploads statistical features extracted from raw vibration data to remote cloud server and we can visualize the data points through an interactive dashboard in real-time. We then explore different machine learning based algorithms to classify different structures based on the collected vibration data, which gives an accuracy up to 91\%. To improve the accuracy, we build a Deep Neural Network and tune its hyperparameters. Accordingly, we achieve an accuracy up to  97\%. Finally, We derive a linear model relating mean \textit{amplitudes} of vibration and floor heights of a building. 

One fundamental paradigm shift realized in our study is that 
we explore the \textit{time} domain of vibration while analyzing the vibrations generated by the different civil structures, as only this domain exhibits considerable values in the case of using low-cost vibration (piezoelectric) sensors. This clearly differs from the existing research studies, which explore the \textit{frequency} domain of vibration while analyzing the vibrations generated by civil structures, as \textit{frequency} domain exhibits considerable values in the case of using high-cost vibration sensors. To the best of our knowledge, we are the first to reveal this finding.

\setlength{\parindent}{3ex} Moving forward, there are several important directions we plan to investigate in future. Examples include - (1) tuning the time series window size, which is considered as eight seconds in this paper, (2) evaluating the power consumption of the system to confirm long-term energy-efficiency, (3) exploring the theoretical background of relationship between mean \textit{amplitudes} of vibration and associated building heights, and (4) exploring specific applications of proposed sensing module in structural health monitoring.






\bibliographystyle{ACM-Reference-Format}
\bibliography{bibfile}


\begin{thebibliography}{48}


\ifx \showCODEN    \undefined \def \showCODEN     #1{\unskip}     \fi
\ifx \showDOI      \undefined \def \showDOI       #1{#1}\fi
\ifx \showISBNx    \undefined \def \showISBNx     #1{\unskip}     \fi
\ifx \showISBNxiii \undefined \def \showISBNxiii  #1{\unskip}     \fi
\ifx \showISSN     \undefined \def \showISSN      #1{\unskip}     \fi
\ifx \showLCCN     \undefined \def \showLCCN      #1{\unskip}     \fi
\ifx \shownote     \undefined \def \shownote      #1{#1}          \fi
\ifx \showarticletitle \undefined \def \showarticletitle #1{#1}   \fi
\ifx \showURL      \undefined \def \showURL       {\relax}        \fi
\providecommand\bibfield[2]{#2}
\providecommand\bibinfo[2]{#2}
\providecommand\natexlab[1]{#1}
\providecommand\showeprint[2][]{arXiv:#2}

\bibitem[\protect\citeauthoryear{Ahmed, El~Sayed, Gadsden, Tjong, and
  Habibi}{Ahmed et~al\mbox{.}}{2014}]%
        {ahmed2014automotive}
\bibfield{author}{\bibinfo{person}{Ryan Ahmed}, \bibinfo{person}{Mohammed
  El~Sayed}, \bibinfo{person}{S~Andrew Gadsden}, \bibinfo{person}{Jimi Tjong},
  {and} \bibinfo{person}{Saeid Habibi}.} \bibinfo{year}{2014}\natexlab{}.
\newblock \showarticletitle{Automotive internal-combustion-engine fault
  detection and classification using artificial neural network techniques}.
\newblock \bibinfo{journal}{\emph{IEEE Transactions on vehicular technology}}
  \bibinfo{volume}{64}, \bibinfo{number}{1} (\bibinfo{year}{2014}),
  \bibinfo{pages}{21--33}.
\newblock


\bibitem[\protect\citeauthoryear{Al-Ayyoub, Nuseir, Alsmearat, Jararweh, and
  Gupta}{Al-Ayyoub et~al\mbox{.}}{2018}]%
        {al2018deep}
\bibfield{author}{\bibinfo{person}{Mahmoud Al-Ayyoub}, \bibinfo{person}{Aya
  Nuseir}, \bibinfo{person}{Kholoud Alsmearat}, \bibinfo{person}{Yaser
  Jararweh}, {and} \bibinfo{person}{Brij Gupta}.}
  \bibinfo{year}{2018}\natexlab{}.
\newblock \showarticletitle{Deep learning for Arabic NLP: A survey}.
\newblock \bibinfo{journal}{\emph{Journal of computational science}}
  \bibinfo{volume}{26} (\bibinfo{year}{2018}), \bibinfo{pages}{522--531}.
\newblock


\bibitem[\protect\citeauthoryear{Bachmann, Ammann, Deischl, Eisenmann, Floegl,
  Hirsch, Klein, Lande, Mahrenholtz, Natke, et~al\mbox{.}}{Bachmann
  et~al\mbox{.}}{2012}]%
        {bachmann2012vibration}
\bibfield{author}{\bibinfo{person}{Hugo Bachmann}, \bibinfo{person}{Walter~J
  Ammann}, \bibinfo{person}{Florian Deischl}, \bibinfo{person}{Josef
  Eisenmann}, \bibinfo{person}{Ingomar Floegl}, \bibinfo{person}{Gerhard~H
  Hirsch}, \bibinfo{person}{G{\"u}nter~K Klein}, \bibinfo{person}{G{\"o}ran~J
  Lande}, \bibinfo{person}{Oskar Mahrenholtz}, \bibinfo{person}{Hans~G Natke},
  {et~al\mbox{.}}} \bibinfo{year}{2012}\natexlab{}.
\newblock \bibinfo{booktitle}{\emph{Vibration problems in structures: practical
  guidelines}}.
\newblock \bibinfo{publisher}{Birkh{\"a}user}.
\newblock


\bibitem[\protect\citeauthoryear{Bangladesh National Building~Code}{Bangladesh
  National Building~Code}{2017a}]%
        {bnbc1}
\bibfield{author}{\bibinfo{person}{BNBC Bangladesh National Building~Code}.}
  \bibinfo{year}{2017}\natexlab{a}.
\newblock \bibinfo{booktitle}{\emph{Definition and general requirements. Part 6
  Chap. 1 Section 3.4}}.
\newblock
\urldef\tempurl%
\url{http://bsa.com.bd/cms_cpanel/upload/pdf_file_upload__1540152875.pdf}
\showURL{%
\tempurl}


\bibitem[\protect\citeauthoryear{Bangladesh National Building~Code}{Bangladesh
  National Building~Code}{2017b}]%
        {bnbc2}
\bibfield{author}{\bibinfo{person}{BNBC Bangladesh National Building~Code}.}
  \bibinfo{year}{2017}\natexlab{b}.
\newblock \bibinfo{booktitle}{\emph{Steel Structures. Part 6 Chap. 10 Section
  12.5}}.
\newblock
\urldef\tempurl%
\url{http://bsa.com.bd/cms_cpanel/upload/pdf_file_upload__1540152875.pdf}
\showURL{%
\tempurl}


\bibitem[\protect\citeauthoryear{Barbosa, Ferreira, Pereira, Magalh{\~a}es, and
  Barbosa}{Barbosa et~al\mbox{.}}{2016}]%
        {barbosa2016fault}
\bibfield{author}{\bibinfo{person}{T{\'a}ssio~S Barbosa},
  \bibinfo{person}{Danton~D Ferreira}, \bibinfo{person}{Daniel~A Pereira},
  \bibinfo{person}{Ricardo~R Magalh{\~a}es}, {and} \bibinfo{person}{Bruno~HG
  Barbosa}.} \bibinfo{year}{2016}\natexlab{}.
\newblock \showarticletitle{Fault detection and classification in cantilever
  beams through vibration signal analysis and higher-order statistics}.
\newblock \bibinfo{journal}{\emph{Journal of Control, Automation and Electrical
  Systems}} \bibinfo{volume}{27}, \bibinfo{number}{5} (\bibinfo{year}{2016}),
  \bibinfo{pages}{535--541}.
\newblock


\bibitem[\protect\citeauthoryear{Berlin and Van~Laerhoven}{Berlin and
  Van~Laerhoven}{2013}]%
        {berlin2013sensor}
\bibfield{author}{\bibinfo{person}{Eugen Berlin} {and} \bibinfo{person}{Kristof
  Van~Laerhoven}.} \bibinfo{year}{2013}\natexlab{}.
\newblock \showarticletitle{Sensor networks for railway monitoring: Detecting
  trains from their distributed vibration footprints}. In
  \bibinfo{booktitle}{\emph{2013 IEEE International Conference on Distributed
  Computing in Sensor Systems}}. IEEE, \bibinfo{pages}{80--87}.
\newblock


\bibitem[\protect\citeauthoryear{Bindi, Petrovic, Karapetrou, Manakou,
  Boxberger, Raptakis, Pitilakis, and Parolai}{Bindi et~al\mbox{.}}{2015}]%
        {bindi2015seismic}
\bibfield{author}{\bibinfo{person}{Dino Bindi}, \bibinfo{person}{Bojana
  Petrovic}, \bibinfo{person}{S Karapetrou}, \bibinfo{person}{M Manakou},
  \bibinfo{person}{Tobias Boxberger}, \bibinfo{person}{Dimitrios Raptakis},
  \bibinfo{person}{KD Pitilakis}, {and} \bibinfo{person}{Stefano Parolai}.}
  \bibinfo{year}{2015}\natexlab{}.
\newblock \showarticletitle{Seismic response of an 8-story RC-building from
  ambient vibration analysis}.
\newblock \bibinfo{journal}{\emph{Bulletin of Earthquake Engineering}}
  \bibinfo{volume}{13}, \bibinfo{number}{7} (\bibinfo{year}{2015}),
  \bibinfo{pages}{2095--2120}.
\newblock


\bibitem[\protect\citeauthoryear{Brain}{Brain}{2015}]%
        {tensorflow}
\bibfield{author}{\bibinfo{person}{Google Brain}.}
  \bibinfo{year}{2015}\natexlab{}.
\newblock \bibinfo{booktitle}{\emph{Tensorflow}}.
\newblock
\urldef\tempurl%
\url{https://www.tensorflow.org/}
\showURL{%
\tempurl}


\bibitem[\protect\citeauthoryear{Chollet}{Chollet}{2015}]%
        {keras}
\bibfield{author}{\bibinfo{person}{François Chollet}.}
  \bibinfo{year}{2015}\natexlab{}.
\newblock \bibinfo{booktitle}{\emph{Keras}}.
\newblock
\urldef\tempurl%
\url{https://keras.io/}
\showURL{%
\tempurl}


\bibitem[\protect\citeauthoryear{Essiet, Sun, and Wang}{Essiet
  et~al\mbox{.}}{2019}]%
        {essiet2019big}
\bibfield{author}{\bibinfo{person}{Ima Essiet}, \bibinfo{person}{Yanxia Sun},
  {and} \bibinfo{person}{Zenghui Wang}.} \bibinfo{year}{2019}\natexlab{}.
\newblock \showarticletitle{Big data analysis for gas sensor using
  convolutional neural network and ensemble of evolutionary algorithms}.
\newblock \bibinfo{journal}{\emph{Procedia Manufacturing}}
  \bibinfo{volume}{35} (\bibinfo{year}{2019}), \bibinfo{pages}{629--634}.
\newblock


\bibitem[\protect\citeauthoryear{Fakhrulddin, Fei, and Li}{Fakhrulddin
  et~al\mbox{.}}{2017}]%
        {fakhrulddin2017convolutional}
\bibfield{author}{\bibinfo{person}{Ali~Haider Fakhrulddin},
  \bibinfo{person}{Xiang Fei}, {and} \bibinfo{person}{Hanchao Li}.}
  \bibinfo{year}{2017}\natexlab{}.
\newblock \showarticletitle{Convolutional neural networks (CNN) based human
  fall detection on body sensor networks (BSN) sensor data}. In
  \bibinfo{booktitle}{\emph{2017 4th international conference on systems and
  informatics (ICSAI)}}. IEEE, \bibinfo{pages}{1461--1465}.
\newblock


\bibitem[\protect\citeauthoryear{Garrity, Bhattacharyya, Shen, Dawadi, and
  Panja}{Garrity et~al\mbox{.}}{[n.d.]}]%
        {garrityvibration}
\bibfield{author}{\bibinfo{person}{P Garrity}, \bibinfo{person}{S
  Bhattacharyya}, \bibinfo{person}{C Shen}, \bibinfo{person}{D Dawadi}, {and}
  \bibinfo{person}{B Panja}.} \bibinfo{year}{[n.d.]}\natexlab{}.
\newblock \showarticletitle{VIBRATION MONITORING AND ANALYSIS USING A WIRELESS
  SENSOR NETWORK (WSN) TO CLASSIFY VEHICLES}.
\newblock  (\bibinfo{year}{[n.\,d.]}).
\newblock


\bibitem[\protect\citeauthoryear{Goyal and Pabla}{Goyal and Pabla}{2016}]%
        {goyal2016vibration}
\bibfield{author}{\bibinfo{person}{D Goyal} {and} \bibinfo{person}{BS Pabla}.}
  \bibinfo{year}{2016}\natexlab{}.
\newblock \showarticletitle{The vibration monitoring methods and signal
  processing techniques for structural health monitoring: a review}.
\newblock \bibinfo{journal}{\emph{Archives of Computational Methods in
  Engineering}} \bibinfo{volume}{23}, \bibinfo{number}{4}
  (\bibinfo{year}{2016}), \bibinfo{pages}{585--594}.
\newblock


\bibitem[\protect\citeauthoryear{Joshuva and Sugumaran}{Joshuva and
  Sugumaran}{2019}]%
        {joshuva2019selection}
\bibfield{author}{\bibinfo{person}{A Joshuva} {and} \bibinfo{person}{V
  Sugumaran}.} \bibinfo{year}{2019}\natexlab{}.
\newblock \showarticletitle{Selection of a meta classifier-data model for
  classifying wind turbine blade fault conditions using histogram features and
  vibration signals: a data-mining study}.
\newblock \bibinfo{journal}{\emph{Progress in Industrial Ecology, an
  International Journal}} \bibinfo{volume}{13}, \bibinfo{number}{3}
  (\bibinfo{year}{2019}), \bibinfo{pages}{232--251}.
\newblock


\bibitem[\protect\citeauthoryear{Junior, Jacques, G{\"u}{\c{c}}l{\"u}t{\"u}rk,
  P{\'e}rez, G{\"u}{\c{c}}l{\"u}, Andujar, Bar{\'o}, Escalante, Guyon, van
  Gerven, et~al\mbox{.}}{Junior et~al\mbox{.}}{2018}]%
        {junior2018first}
\bibfield{author}{\bibinfo{person}{JCSJ Junior}, \bibinfo{person}{C Jacques},
  \bibinfo{person}{Ya{\u{g}}mur G{\"u}{\c{c}}l{\"u}t{\"u}rk},
  \bibinfo{person}{Marc P{\'e}rez}, \bibinfo{person}{Umut G{\"u}{\c{c}}l{\"u}},
  \bibinfo{person}{Carlos Andujar}, \bibinfo{person}{Xavier Bar{\'o}},
  \bibinfo{person}{Hugo~Jair Escalante}, \bibinfo{person}{Isabelle Guyon},
  \bibinfo{person}{MA van Gerven}, {et~al\mbox{.}}}
  \bibinfo{year}{2018}\natexlab{}.
\newblock \showarticletitle{First impressions: A survey on computer
  vision-based apparent personality trait analysis}.
\newblock \bibinfo{journal}{\emph{arXiv preprint arXiv:1804.08046}}
  (\bibinfo{year}{2018}).
\newblock


\bibitem[\protect\citeauthoryear{Kamal, Ahmed, Toha, Islam, and Al~Islam}{Kamal
  et~al\mbox{.}}{2019}]%
        {kamal2019intelligent}
\bibfield{author}{\bibinfo{person}{Uday Kamal}, \bibinfo{person}{Shamir Ahmed},
  \bibinfo{person}{Tarik~Reza Toha}, \bibinfo{person}{Nafisa Islam}, {and}
  \bibinfo{person}{ABM~Alim Al~Islam}.} \bibinfo{year}{2019}\natexlab{}.
\newblock \showarticletitle{Intelligent Human Counting through Environmental
  Sensing in Closed Indoor Settings}.
\newblock \bibinfo{journal}{\emph{Mobile Networks and Applications}}
  (\bibinfo{year}{2019}), \bibinfo{pages}{1--17}.
\newblock


\bibitem[\protect\citeauthoryear{keras}{keras}{2020}]%
        {adam}
\bibfield{author}{\bibinfo{person}{keras}.} \bibinfo{year}{2020}\natexlab{}.
\newblock \bibinfo{booktitle}{\emph{Adam optimizer}}.
\newblock
\urldef\tempurl%
\url{https://keras.io/api/optimizers/adam/}
\showURL{%
\tempurl}


\bibitem[\protect\citeauthoryear{Khan, Sohail, Zahoora, and Qureshi}{Khan
  et~al\mbox{.}}{2019}]%
        {khan2019survey}
\bibfield{author}{\bibinfo{person}{Asifullah Khan}, \bibinfo{person}{Anabia
  Sohail}, \bibinfo{person}{Umme Zahoora}, {and} \bibinfo{person}{Aqsa~Saeed
  Qureshi}.} \bibinfo{year}{2019}\natexlab{}.
\newblock \showarticletitle{A survey of the recent architectures of deep
  convolutional neural networks}.
\newblock \bibinfo{journal}{\emph{Artificial Intelligence Review}}
  (\bibinfo{year}{2019}), \bibinfo{pages}{1--62}.
\newblock


\bibitem[\protect\citeauthoryear{K{\"u}{\c{c}}{\"u}kbay, Sert, and
  Yazici}{K{\"u}{\c{c}}{\"u}kbay et~al\mbox{.}}{2017}]%
        {kuccukbay2017use}
\bibfield{author}{\bibinfo{person}{Selver~Ezgi K{\"u}{\c{c}}{\"u}kbay},
  \bibinfo{person}{Mustafa Sert}, {and} \bibinfo{person}{Adnan Yazici}.}
  \bibinfo{year}{2017}\natexlab{}.
\newblock \showarticletitle{Use of acoustic and vibration sensor data to detect
  objects in surveillance wireless sensor networks}. In
  \bibinfo{booktitle}{\emph{2017 21st International Conference on Control
  Systems and Computer Science (CSCS)}}. IEEE, \bibinfo{pages}{207--212}.
\newblock


\bibitem[\protect\citeauthoryear{Lagomarsino}{Lagomarsino}{1993}]%
        {lagomarsino1993forecast}
\bibfield{author}{\bibinfo{person}{Sergio Lagomarsino}.}
  \bibinfo{year}{1993}\natexlab{}.
\newblock \showarticletitle{Forecast models for damping and vibration periods
  of buildings}.
\newblock \bibinfo{journal}{\emph{Journal of Wind Engineering and Industrial
  Aerodynamics}} \bibinfo{volume}{48}, \bibinfo{number}{2-3}
  (\bibinfo{year}{1993}), \bibinfo{pages}{221--239}.
\newblock


\bibitem[\protect\citeauthoryear{Learning}{Learning}{2020}]%
        {epoch}
\bibfield{author}{\bibinfo{person}{Deep Learning}.}
  \bibinfo{year}{2020}\natexlab{}.
\newblock \bibinfo{booktitle}{\emph{Epochs}}.
\newblock
\urldef\tempurl%
\url{https://towardsdatascience.com/epoch-vs-iterations-vs-batch-size-4dfb9c7ce9c9}
\showURL{%
\tempurl}


\bibitem[\protect\citeauthoryear{Li, Wu, Liang, Xiao, and Wong}{Li
  et~al\mbox{.}}{2004}]%
        {li2004full}
\bibfield{author}{\bibinfo{person}{QS Li}, \bibinfo{person}{JR Wu},
  \bibinfo{person}{SG Liang}, \bibinfo{person}{YQ Xiao}, {and}
  \bibinfo{person}{CK Wong}.} \bibinfo{year}{2004}\natexlab{}.
\newblock \showarticletitle{Full-scale measurements and numerical evaluation of
  wind-induced vibration of a 63-story reinforced concrete tall building}.
\newblock \bibinfo{journal}{\emph{Engineering structures}}
  \bibinfo{volume}{26}, \bibinfo{number}{12} (\bibinfo{year}{2004}),
  \bibinfo{pages}{1779--1794}.
\newblock


\bibitem[\protect\citeauthoryear{Li, Yang, Wong, and Jeary}{Li
  et~al\mbox{.}}{2003}]%
        {li2003effect}
\bibfield{author}{\bibinfo{person}{QS Li}, \bibinfo{person}{Ke Yang},
  \bibinfo{person}{CK Wong}, {and} \bibinfo{person}{AP Jeary}.}
  \bibinfo{year}{2003}\natexlab{}.
\newblock \showarticletitle{The effect of amplitude-dependent damping on
  wind-induced vibrations of a super tall building}.
\newblock \bibinfo{journal}{\emph{Journal of Wind Engineering and Industrial
  Aerodynamics}} \bibinfo{volume}{91}, \bibinfo{number}{9}
  (\bibinfo{year}{2003}), \bibinfo{pages}{1175--1198}.
\newblock


\bibitem[\protect\citeauthoryear{LLC}{LLC}{2010}]%
        {kaggle}
\bibfield{author}{\bibinfo{person}{Google LLC}.}
  \bibinfo{year}{2010}\natexlab{}.
\newblock \bibinfo{booktitle}{\emph{Kaggle}}.
\newblock
\urldef\tempurl%
\url{https://www.kaggle.com/}
\showURL{%
\tempurl}


\bibitem[\protect\citeauthoryear{Magalh{\~a}es, Cunha, and
  Caetano}{Magalh{\~a}es et~al\mbox{.}}{2012}]%
        {magalhaes2012vibration}
\bibfield{author}{\bibinfo{person}{Filipe Magalh{\~a}es},
  \bibinfo{person}{{\'A}lvaro Cunha}, {and} \bibinfo{person}{Elsa Caetano}.}
  \bibinfo{year}{2012}\natexlab{}.
\newblock \showarticletitle{Vibration based structural health monitoring of an
  arch bridge: from automated OMA to damage detection}.
\newblock \bibinfo{journal}{\emph{Mechanical Systems and Signal Processing}}
  \bibinfo{volume}{28} (\bibinfo{year}{2012}), \bibinfo{pages}{212--228}.
\newblock


\bibitem[\protect\citeauthoryear{MATLAB}{MATLAB}{2020}]%
        {global_average_pooling}
\bibfield{author}{\bibinfo{person}{MATLAB}.} \bibinfo{year}{2020}\natexlab{}.
\newblock \bibinfo{booktitle}{\emph{Global Average Pooling}}.
\newblock
\urldef\tempurl%
\url{https://www.mathworks.com/help/deeplearning/ref/nnet.cnn.layer.globalaveragepooling2dlayer.html#:~:text=Since%20global%20pooling%20layers%20have,layer%20with%20a%20globalMaxPooling2dLayer%20instead.}
\showURL{%
\tempurl}


\bibitem[\protect\citeauthoryear{Michel, Gu{\'e}guen, El~Arem, Mazars, and
  Kotronis}{Michel et~al\mbox{.}}{2010}]%
        {michel2010full}
\bibfield{author}{\bibinfo{person}{Clotaire Michel}, \bibinfo{person}{Philippe
  Gu{\'e}guen}, \bibinfo{person}{Saber El~Arem}, \bibinfo{person}{Jacky
  Mazars}, {and} \bibinfo{person}{Panagiotis Kotronis}.}
  \bibinfo{year}{2010}\natexlab{}.
\newblock \showarticletitle{Full-scale dynamic response of an RC building under
  weak seismic motions using earthquake recordings, ambient vibrations and
  modelling}.
\newblock \bibinfo{journal}{\emph{Earthquake Engineering \& Structural
  Dynamics}} \bibinfo{volume}{39}, \bibinfo{number}{4} (\bibinfo{year}{2010}),
  \bibinfo{pages}{419--441}.
\newblock


\bibitem[\protect\citeauthoryear{Numpy}{Numpy}{2020}]%
        {polyfit}
\bibfield{author}{\bibinfo{person}{Numpy}.} \bibinfo{year}{2020}\natexlab{}.
\newblock \bibinfo{booktitle}{\emph{Polyfit}}.
\newblock
\urldef\tempurl%
\url{https://numpy.org/doc/stable/reference/generated/numpy.polyfit.html}
\showURL{%
\tempurl}


\bibitem[\protect\citeauthoryear{Pan, Goh, and Megawati}{Pan
  et~al\mbox{.}}{2014}]%
        {pan2014empirical}
\bibfield{author}{\bibinfo{person}{Tso-Chien Pan}, \bibinfo{person}{Key~Seng
  Goh}, {and} \bibinfo{person}{Kusnowidjaja Megawati}.}
  \bibinfo{year}{2014}\natexlab{}.
\newblock \showarticletitle{Empirical relationships between natural vibration
  period and height of buildings in Singapore}.
\newblock \bibinfo{journal}{\emph{Earthquake engineering \& structural
  dynamics}} \bibinfo{volume}{43}, \bibinfo{number}{3} (\bibinfo{year}{2014}),
  \bibinfo{pages}{449--465}.
\newblock


\bibitem[\protect\citeauthoryear{Python}{Python}{2020a}]%
        {numpy}
\bibfield{author}{\bibinfo{person}{Python}.} \bibinfo{year}{2020}\natexlab{a}.
\newblock \bibinfo{booktitle}{\emph{Numpy statistical functions}}.
\newblock
\urldef\tempurl%
\url{https://numpy.org/doc/stable/reference/routines.statistics.html}
\showURL{%
\tempurl}


\bibitem[\protect\citeauthoryear{Python}{Python}{2020b}]%
        {scipy}
\bibfield{author}{\bibinfo{person}{Python}.} \bibinfo{year}{2020}\natexlab{b}.
\newblock \bibinfo{booktitle}{\emph{Scipy statistical functions}}.
\newblock
\urldef\tempurl%
\url{https://docs.scipy.org/doc/scipy/reference/stats.html}
\showURL{%
\tempurl}


\bibitem[\protect\citeauthoryear{Ravi, Wong, Lo, and Yang}{Ravi
  et~al\mbox{.}}{2016}]%
        {ravi2016deep}
\bibfield{author}{\bibinfo{person}{Daniele Ravi}, \bibinfo{person}{Charence
  Wong}, \bibinfo{person}{Benny Lo}, {and} \bibinfo{person}{Guang-Zhong Yang}.}
  \bibinfo{year}{2016}\natexlab{}.
\newblock \showarticletitle{A deep learning approach to on-node sensor data
  analytics for mobile or wearable devices}.
\newblock \bibinfo{journal}{\emph{IEEE journal of biomedical and health
  informatics}} \bibinfo{volume}{21}, \bibinfo{number}{1}
  (\bibinfo{year}{2016}), \bibinfo{pages}{56--64}.
\newblock


\bibitem[\protect\citeauthoryear{Richman and Deadrick}{Richman and
  Deadrick}{2013}]%
        {richman2013seismic}
\bibfield{author}{\bibinfo{person}{Michael~S Richman} {and}
  \bibinfo{person}{Douglas~S Deadrick}.} \bibinfo{year}{2013}\natexlab{}.
\newblock \bibinfo{title}{Seismic method for vehicle detection and vehicle
  weight classification}.
\newblock
\newblock
\newblock
\shownote{US Patent 8,405,524.}


\bibitem[\protect\citeauthoryear{Rivas, Wunderlich, and Heinen}{Rivas
  et~al\mbox{.}}{2016}]%
        {rivas2016road}
\bibfield{author}{\bibinfo{person}{Javier Rivas}, \bibinfo{person}{Ralf
  Wunderlich}, {and} \bibinfo{person}{Stefan~J Heinen}.}
  \bibinfo{year}{2016}\natexlab{}.
\newblock \showarticletitle{Road vibrations as a source to detect the presence
  and speed of vehicles}.
\newblock \bibinfo{journal}{\emph{IEEE Sensors Journal}} \bibinfo{volume}{17},
  \bibinfo{number}{2} (\bibinfo{year}{2016}), \bibinfo{pages}{377--385}.
\newblock


\bibitem[\protect\citeauthoryear{scikit learn}{scikit learn}{2020a}]%
        {fa}
\bibfield{author}{\bibinfo{person}{scikit learn}.}
  \bibinfo{year}{2020}\natexlab{a}.
\newblock \bibinfo{booktitle}{\emph{Factor Analysis (FA)}}.
\newblock
\urldef\tempurl%
\url{https://scikit-learn.org/stable/modules/generated/sklearn.decomposition.FactorAnalysis.html}
\showURL{%
\tempurl}


\bibitem[\protect\citeauthoryear{scikit learn}{scikit learn}{2020b}]%
        {scikit-learn}
\bibfield{author}{\bibinfo{person}{scikit learn}.}
  \bibinfo{year}{2020}\natexlab{b}.
\newblock \bibinfo{booktitle}{\emph{Machine Learning in Python}}.
\newblock
\urldef\tempurl%
\url{https://scikit-learn.org/stable/}
\showURL{%
\tempurl}


\bibitem[\protect\citeauthoryear{scikit learn}{scikit learn}{2020c}]%
        {pca}
\bibfield{author}{\bibinfo{person}{scikit learn}.}
  \bibinfo{year}{2020}\natexlab{c}.
\newblock \bibinfo{booktitle}{\emph{Principle Component Analysis (PCA)}}.
\newblock
\urldef\tempurl%
\url{https://scikit-learn.org/stable/modules/generated/sklearn.decomposition.PCA.html}
\showURL{%
\tempurl}


\bibitem[\protect\citeauthoryear{scikit learn}{scikit learn}{2020d}]%
        {tsne}
\bibfield{author}{\bibinfo{person}{scikit learn}.}
  \bibinfo{year}{2020}\natexlab{d}.
\newblock \bibinfo{booktitle}{\emph{t-distributed Stochastic Neighbor Embedding
  (t-SNE)}}.
\newblock
\urldef\tempurl%
\url{https://scikit-learn.org/stable/modules/generated/sklearn.manifold.TSNE.html}
\showURL{%
\tempurl}


\bibitem[\protect\citeauthoryear{Scipy}{Scipy}{2019}]%
        {pearson}
\bibfield{author}{\bibinfo{person}{Scipy}.} \bibinfo{year}{2019}\natexlab{}.
\newblock \bibinfo{booktitle}{\emph{Statistics in Python}}.
\newblock
\urldef\tempurl%
\url{https://scipy-lectures.org/packages/statistics/index.html}
\showURL{%
\tempurl}


\bibitem[\protect\citeauthoryear{Scipy}{Scipy}{2020}]%
        {fft}
\bibfield{author}{\bibinfo{person}{Scipy}.} \bibinfo{year}{2020}\natexlab{}.
\newblock \bibinfo{booktitle}{\emph{Fast Fourier Transform}}.
\newblock
\urldef\tempurl%
\url{https://docs.scipy.org/doc/scipy/reference/tutorial/fft.html}
\showURL{%
\tempurl}


\bibitem[\protect\citeauthoryear{Sigmund, Shelley, Bauer, and Heitkamp}{Sigmund
  et~al\mbox{.}}{2012}]%
        {sigmund2012analysis}
\bibfield{author}{\bibinfo{person}{Kevin~J Sigmund}, \bibinfo{person}{Stuart~J
  Shelley}, \bibinfo{person}{Mitchell Bauer}, {and} \bibinfo{person}{Frederick
  Heitkamp}.} \bibinfo{year}{2012}\natexlab{}.
\newblock \showarticletitle{Analysis of vehicle vibration sources for automatic
  differentiation between gas and diesel piston engines}. In
  \bibinfo{booktitle}{\emph{Automatic Target Recognition XXII}},
  Vol.~\bibinfo{volume}{8391}. International Society for Optics and Photonics,
  \bibinfo{pages}{839109}.
\newblock


\bibitem[\protect\citeauthoryear{Sindagi and Patel}{Sindagi and Patel}{2018}]%
        {sindagi2018survey}
\bibfield{author}{\bibinfo{person}{Vishwanath~A Sindagi} {and}
  \bibinfo{person}{Vishal~M Patel}.} \bibinfo{year}{2018}\natexlab{}.
\newblock \showarticletitle{A survey of recent advances in cnn-based single
  image crowd counting and density estimation}.
\newblock \bibinfo{journal}{\emph{Pattern Recognition Letters}}
  \bibinfo{volume}{107} (\bibinfo{year}{2018}), \bibinfo{pages}{3--16}.
\newblock


\bibitem[\protect\citeauthoryear{sklearn}{sklearn}{2020}]%
        {crossval}
\bibfield{author}{\bibinfo{person}{sklearn}.} \bibinfo{year}{2020}\natexlab{}.
\newblock \bibinfo{booktitle}{\emph{K-fold cross-validation}}.
\newblock
\urldef\tempurl%
\url{https://scikit-learn.org/stable/modules/generated/sklearn.model_selection.KFold.html}
\showURL{%
\tempurl}


\bibitem[\protect\citeauthoryear{Sparkfun}{Sparkfun}{[n.d.]a}]%
        {arduino_mega}
\bibfield{author}{\bibinfo{person}{Sparkfun}.}
  \bibinfo{year}{[n.d.]}\natexlab{a}.
\newblock \bibinfo{booktitle}{\emph{Arduino Mega 2560}}.
\newblock
\urldef\tempurl%
\url{https://www.sparkfun.com/products/11061}
\showURL{%
\tempurl}


\bibitem[\protect\citeauthoryear{Sparkfun}{Sparkfun}{[n.d.]b}]%
        {piezo}
\bibfield{author}{\bibinfo{person}{Sparkfun}.}
  \bibinfo{year}{[n.d.]}\natexlab{b}.
\newblock \bibinfo{booktitle}{\emph{Piezo Element}}.
\newblock
\urldef\tempurl%
\url{https://www.sparkfun.com/products/10293}
\showURL{%
\tempurl}


\bibitem[\protect\citeauthoryear{Techshop}{Techshop}{2017}]%
        {sim900a}
\bibfield{author}{\bibinfo{person}{Techshop}.} \bibinfo{year}{2017}\natexlab{}.
\newblock \bibinfo{booktitle}{\emph{SIM900A}}.
\newblock
\urldef\tempurl%
\url{https://www.techshopbd.com/product-categories/eval-board/2041/sim900a-kit-techshop-bangladesh}
\showURL{%
\tempurl}


\bibitem[\protect\citeauthoryear{Wang, Chen, Hao, Peng, and Hu}{Wang
  et~al\mbox{.}}{2019}]%
        {wang2019deep}
\bibfield{author}{\bibinfo{person}{Jindong Wang}, \bibinfo{person}{Yiqiang
  Chen}, \bibinfo{person}{Shuji Hao}, \bibinfo{person}{Xiaohui Peng}, {and}
  \bibinfo{person}{Lisha Hu}.} \bibinfo{year}{2019}\natexlab{}.
\newblock \showarticletitle{Deep learning for sensor-based activity
  recognition: A survey}.
\newblock \bibinfo{journal}{\emph{Pattern Recognition Letters}}
  \bibinfo{volume}{119} (\bibinfo{year}{2019}), \bibinfo{pages}{3--11}.
\newblock


\end{thebibliography}

\appendix

\end{document}